\documentclass[aps,superscriptaddress,showpacs,nofootinbib, 10pt]{revtex4-2}
\usepackage{mathrsfs}
\usepackage{bigints}
\usepackage{latexsym,bm}
\usepackage{graphicx}
\usepackage{indentfirst}
\usepackage{slashed}
\usepackage{amssymb}
\usepackage{amsmath}
\usepackage{bbm}
\usepackage{color}
\usepackage[dvipsnames]{xcolor}
\usepackage{epsfig}
\usepackage[titletoc]{appendix}
\usepackage{multirow}%
\usepackage{rotating}
\usepackage{epstopdf}
\usepackage{extarrows}
\usepackage{tabularx}
\usepackage{soul} 
\usepackage{xcolor}
\usepackage{lipsum}
\usepackage[colorlinks=true,allcolors=BlueViolet,hyperfootnotes=false]{hyperref}

\usepackage{ulem}

\usepackage[utf8]{inputenc}

\newcommand{\be}{\begin{equation}} \newcommand{\ee}{\end{equation}}
\newcommand{\ba}{\begin{array}{c}} \newcommand{\ea}{\end{array}}
\newcommand{\bea}{\begin{eqnarray}} \newcommand{\eea}{\end{eqnarray}}
\newcommand{\beas}{\begin{eqnarray*}} \newcommand{\eeas}{\end{eqnarray*}}

\newcommand{\no}{\nonumber\\}

\begin{document}
\title{\Large  Inclusive quasielastic (anti-)neutrino nucleus scattering within the Standard Model and beyond }
\author{E. Hern\'andez}
\affiliation{Departamento de F\'\i sica Fundamental 
  e IUFFyM,\\ Universidad de Salamanca, E-37008 Salamanca, Spain}
\author{J. Nieves}
\affiliation{Instituto de F\'{\i}sica Corpuscular (centro mixto CSIC-UV), Institutos de Investigaci\'on de Paterna,
Apartado 22085, 46071, Valencia, Spain}
\author{J. E. Sobczyk}
\affiliation{Department of Physics, Chalmers University of Technology, SE-412 96 G\"oteborg, Sweden}
\affiliation{Institut f\"ur Kernphysik and PRISMA+ Cluster of 
Excellence, Johannes Gutenberg-Universit\"at Mainz, 55128 Mainz, Germany}

\date{\today}

\begin{abstract}
We derive fully general expressions for the inclusive (anti-)neutrino-induced nuclear quasielastic production of a strange or  charmed baryon $Y$,  considering all dimension-six new physics operators relevant to the semileptonic $q\to q'\ell \nu_\ell$ transition with  both left- and right-handed neutrino fields. We illustrate the formalism by applying it to the $\Lambda_c$ production in neutrino-nucleus scattering. 
The nuclear response is computed using a state-of-the-art {\it ab initio}  spectral function, based on realistic nuclear many-body wavefunctions obtained via coupled-cluster method and a nuclear Hamiltonian derived from chiral effective field theory.
We also highlight the role played by the final state interactions between the produced $Y$ hyperon and the residual nuclear system, which can obscure potential new physics signals in this (anti-)neutrino–nucleus reaction. Our results improve upon previous studies that neglected nuclear corrections and uncertainties—effects we show to be comparable to, or even larger than, those expected from new physics.


\end{abstract}

\maketitle
\section{Introduction}
Despite its remarkable success in describing a wide range of experimental observations, the Standard Model (SM) is widely regarded as a low-energy limit of a more fundamental theory. This perspective is supported not only by theoretical considerations (see, e.g., Chapter 10 of Ref.~\cite{langacker:2017}) but also by experimental results that remain unexplained within the SM. One key feature of the SM is lepton flavor universality —the principle that all three lepton families couple identically with the $W$ and $Z$ gauge bosons. It is precisely this feature that has been called into question by recent experimental anomalies.
The most compelling indication of possible  lepton flavor universality violation comes from the ratios 
${\cal R}_{D^{(*)}}=\Gamma(\bar B\to D^{(*)}\tau^-\bar\nu_\tau)/
\Gamma(\bar B\to D^{(*)}\ell^-\bar\nu_\ell)$
for which the HFLAV collaboration~\cite{HeavyFlavorAveragingGroupHFLAV:2024ctg}
has reported experimental averages that deviate from SM predictions at the level of $3.1\,\sigma$.
A related observable, ${\cal R}_{J/\psi}=
  \Gamma(\bar B_c\to J/\psi\tau^-\bar\nu_\tau)/\Gamma(\bar B_c\to
   J/\psi\mu^-\bar\nu_\mu)$, measured by the   LHCb 
   Collaboration~\cite{LHCb:2017vlu} also shows a tension with SM expectations, with a deviation of approximately $1.8\,\sigma$
   ~\cite{Anisimov:1998uk,Ivanov:2006ni,
Hernandez:2006gt,Huang:2007kb,Wang:2008xt,Wen-Fei:2013uea, Watanabe:2017mip, Issadykov:2018myx,Tran:2018kuv,
Hu:2019qcn,Leljak:2019eyw,Azizi:2019aaf,Wang:2018duy}. 
However, not all measurements point to deviations from the SM. For instance, the ratio
${\cal R}_{\Lambda_c}=
  \Gamma(\Lambda^0_b\to \Lambda^+_c\tau^-\bar\nu_\tau)/\Gamma(\Lambda^0_b\to 
  \Lambda^+_c
  \ell^-\bar\nu_\ell)$
as measured by the LHCb Collaboration~\cite{LHCb:2022piu}, is consistent with SM predictions from Ref.~\cite{Bernlochner:2018bfn}. 
Similarly, the first measurement of the inclusive semileptonic decay ratio
${\cal R}_{X_{\tau/\ell}}=\Gamma(\bar B\to X\tau^-\bar\nu_\tau)/\Gamma(\bar B\to X\ell^-\bar\nu_\ell)$, recently reported by Belle II~\cite{Belle-II:2023aih}, is compatible with the SM predictions from Refs.~\cite{Ligeti:2021six,Rahimi:2022vlv}.
If physics beyond the SM (BSM) is indeed responsible for lepton flavor universality violating effects, one would expect such contributions to manifest across transitions involving multiple quark generations. Nonetheless, the current inconclusive results suggest that any BSM effects at the energy scales probed by present experiments must be moderate. Therefore, achieving high-precision experimental data, alongside equally precise theoretical calculations, is essential for reliably uncovering or constraining possible new physics (NP).

 
A widely used, model-independent approach, to study the relevance of BSM physics effects at  the energy regime accessible to current experiments is 
through an effective field theory analysis.
In this framework, one adopts a phenomenological perspective by systematically incorporating all possible low-energy dimension-six operators mediating $q\leftrightarrow q'$ transitions. These operators are assumed to arise from NP effects at higher energy scales. One of the pioneering works following this strategy is found in Ref.~\cite{Fajfer:2012vx}.
The strengths of these NP-induced operators are parameterized by unknown Wilson coefficients (WCs), which can, in principle, be constrained or extracted through fits to experimental data. As detailed in Sec.~\ref{sec:effhamil}, the effective Lagrangian includes operators corresponding to various interaction types: vector, axial-vector, scalar, pseudoscalar, tensor, and pseudotensor.

For the $d\leftrightarrow c$ quark-level transition, valuable information can be obtained from the study of the ratio
${\cal  R}_{\tau/\mu}=\Gamma(D^+\to\tau^+\nu_\tau)/ \Gamma(D^+\to\mu^+\nu_\mu)$. This observable was first investigated by the CLEO Collaboration~\cite{CLEO:2006jxt}, which reported a statistically non-significant result and provided only an upper bound. A more recent measurement by the BESIII Collaboration~\cite{BESIII:2019vhn} yielded a central value larger than the SM prediction, though still consistent with it within $1\sigma$ uncertainty. Taking this result at face value, several studies have attempted to constrain the corresponding WCs within the effective field theory framework~\cite{Fleischer:2019wlx,Becirevic:2020rzi,Leng:2020fei,Fuentes-Martin:2020lea}.
It is important to note that these purely leptonic decays\footnote{The semileptonic decay to a final-state $\tau$ is kinematically forbidden due to phase-space limitations.} are only sensitive to axial-vector and pseudoscalar WCs. In contrast, vector, scalar, and tensor operators can be probed through complementary processes, such as mono-lepton production in high-energy charged-current Drell–Yan reactions~\cite{Fuentes-Martin:2020lea}.
 
In Ref.~\cite{Kong:2023kkd},  Y.-R. Kong et al. proposed the quasielastic (QE) neutrino scattering process $\nu_\tau n\to\tau^-\Lambda_c$, induced by a $\nu_\tau d\to\tau^- c$ transition at the quark level, as a potential probe to further constrain the WCs associated with a $d\leftrightarrow c$ transition. This reaction could, in principle, be measured at the DUNE far detector, where a fraction of the initial $\nu_\mu$ neutrinos produced at Fermilab will oscillate into $\nu_\tau$ neutrinos. However, the analysis in Ref.~\cite{Kong:2023kkd} is performed at the nucleon level, which is not realistic, since free neutron targets are not available in practice. Nuclear effects are likely to play a significant role, and more importantly, the  uncertainties associated with their treatment may hinder the extraction of any reliable constraints on possible NP WCs—even under optimistic assumptions such as perfect knowledge of the nucleon form factors and extremely high event statistics.

In this work, under the latter conditions,  we revisit the problem and derive general expressions for the inclusive QE reaction 
\begin{equation}
 (\bar \nu_\ell ) 
\nu_\ell A_Z \to (\ell^+)\ell^- Y X   
\end{equation}
where $X$ is a residual nuclear system with $(A-1)$ nucleons, $Y$ is either  a strange or a charmed hyperon and $\ell=e,\, \mu,\,\tau$. We  consider an effective Hamiltonian which is a low energy extension of the SM that includes  all dimension-six semileptonic $q\to q'\ell \nu_\ell$ ($d\to u,c$ and  $u\to d,s$) operators with left- and right-handed neutrino terms. 
All expressions are obtained in terms of NP WCs, and hadron structure functions. The latter are determined by the form factors that describe the matrix elements of the different current operators between the initial $(n,p)$ and final $Y$ hadrons. 

To illustrate the formalism, we apply it to the case of $\Lambda_c$ production in $\nu_\tau$-nucleus QE scattering, as originally proposed in Ref.~\cite{Kong:2023kkd}, but now incorporating a realistic nuclear target and nuclear structure effects through hole spectral functions.
After presenting a general formalism in Sec.~\ref{sec:nucleus}, we assess the impact of nuclear effects in Sec.\ref{sec:results} by comparing predictions obtained using a state-of-the-art spectral function~\cite{Sobczyk:2023mey} with those based on the local Fermi gas (LFG) approximation. 
Additionally, we account for final state interactions involving the outgoing $\Lambda_c$ hyperon. Finally, we evaluate the magnitude of nuclear uncertainties in our calculations relative to the size of the expected NP effects, thereby highlighting the challenges inherent in interpreting such processes as clean probes for BSM physics.

 \section{Effective Hamiltonian}
 \label{sec:effhamil}
 For the $q\,(-e/3)\leftrightarrow q'(2e/3)$ charged-current weak transitions at the quark level  we   shall  consider
\bea
 H^{qq'\ell }_{\rm eff}&=&\frac{4G_F V_{qq'}}{\sqrt2}\sum_\ell\Bigg\{\sum_{\chi,\chi'=L,R}\Big[
 \left(\bar q'\gamma^\mu[\delta_{\chi'L}\delta_{\chi L}+C^V_{qq'\ell\chi'\chi}]
 P_{\chi'}q\right)
 (\bar\ell\gamma_\mu  P_\chi\nu_\ell)
 +C^S_{qq'\ell\chi'\chi}\,(\bar q'
 P_{\chi'}q)
 (\bar\ell  P_\chi\nu_\ell)\Big]\no
 &+&\sum_{\chi=L,R} C^T_{qq'\ell\chi\chi}\left(\bar q'\,\sigma^{\mu\nu}
 P_\chi q\right)
 \left(\bar\ell \sigma_{\mu\nu} P_\chi\nu_\ell\right)\Bigg\},\label{eq:heffqq'tot}
 \eea
where $q(x)$, $q'(x)$, $\ell(x)$ and  $\nu_\ell(x)$ are Dirac fields,
 $P_{\chi=L,R}=\frac12(I+\gamma_5 h_{\chi=L,R})$ with $h_L=-1,h_R=+1$, $G_F$ is the Fermi decay constant and
$V_{qq'}$ the corresponding Cabibbo-Kobayashi-Maskawa matrix element. The above  effective 
 Hamiltonian contains all dimension-six vector, axial-vector, scalar, pseudoscalar, tensor, and pseudotensor transition operators, with both left- and right-handed neutrino fields.  We have explicitly shown only
the terms relevant for neutrino-induced reactions $\nu_\ell q\to q'\ell^-$ and the $q\to q'\ell^-\bar\nu_\ell$ decays.   
For fixed $q,q',\ell$ flavors, the ten (generally complex)  WCs
$C^{V,S}_{qq'\ell\chi'\chi}$ and 
$C^{T}_{qq'\ell\chi\chi}$ (with $\chi,\chi'=L,R$) 
parameterize the strength of possible contributions from BSM physics. Note that for tensor operators, the chiralities of the leptons and quarks have to be the same. This is a consequence of the fact that (we use the convention $\epsilon_{0123}=+1$)
 \be
\sigma^{\alpha\beta}\gamma_5=
 -\frac{i}2\epsilon^{\alpha\beta\rho\lambda}\,\sigma_{\rho\lambda},
\ee
  from where
 \be
\sigma^{\alpha\beta}(I+h_{\chi'}\gamma_5)\otimes\sigma_{\alpha\beta}(I+h_{\chi}\gamma_5)=
(1+h_{\chi'}h_\chi)\sigma^{\alpha\beta}\otimes\sigma_{\alpha\beta}-
(h_{\chi'}+h_\chi)\frac{i}2\epsilon^{\mu\nu}_{\ \ \alpha\beta}\sigma^{\alpha\beta}
\otimes\sigma_{\mu\nu}.
 \ee
 which is zero for $\chi\ne\chi'$.
 
 It is convenient to rewrite $H^{q q'\ell }_{\rm eff}$ as
\bea
 H^{q q'\ell}_{\rm eff}&=&\frac{2G_F V_{qq'}}{\sqrt2}
\sum_\ell \sum_{\chi=L,R}\Big[\left(\bar q'\,[C^V_{qq'\ell\chi} \gamma^\mu+h_\chi 
 C^A_{qq'\ell\chi}
 \gamma^\mu\gamma_5]\,q\right) (\bar \ell\gamma^\mu P_\chi\nu_\ell)+
\left(\bar q'\,[C^S_{qq'\ell\chi} +h_\chi C^P_{qq'\ell\chi}
\gamma_5]\,q\right) (\bar \ell P_\chi\nu_\ell)\no
&+&
C^T_{qq'\ell\chi}\left(\bar q'\, \sigma^{\mu\nu}(I+h_\chi\gamma_5)\,q\right) (\bar \ell 
\sigma_{\mu\nu}P_\chi\nu_\ell)\Big],
\label{eq:heffqq'}
 \eea
with~\cite{Penalva:2021wye}
\bea
&&C^V_{qq'\ell L}=1+C^V_{qq'\ell LL}+C^V_{qq'\ell RL},\ 
C^A_{qq'\ell L}=1+C^V_{qq'\ell LL}-C^V_{qq'\ell RL},\no
&& C^V_{qq'\ell R}=C^V_{qq'\ell LR}+C^V_{qq'\ell RR},\ 
C^A_{qq'\ell R}=C^V_{qq'\ell RR}-C^V_{qq'\ell LR},\no
&&C^S_{qq'\ell L}=C^S_{qq'\ell LL}+C^S_{qq'\ell RL},\ 
C^P_{qq'\ell L}=C^S_{qq'\ell LL}-C^S_{qq'\ell RL},\no
&& C^S_{qq'\ell R}=C^S_{qq'\ell LR}+C^S_{qq'\ell RR},\ 
C^P_{qq'\ell R}=C^S_{qq'\ell RR}-C^S_{qq'\ell LR},\no 
&&C^T_{qq'\ell L}=C^T_{qq'\ell LL},\  C^T_{qq'\ell R}=C^T_{qq'\ell RR}.\label{eq:wcs0}
\eea

Let us now look at the $H^{q q'\bar \ell}_{\rm eff}$ terms that are responsible for the antineutrino-induced 
reactions $\bar\nu_\ell q'\to q\ell^+$ or the  $q'\to q\ell^+\nu_\ell$ decays.
They are obtained from the hermitian conjugate  of the operator $H^{q q'\ell}_{\rm eff}$, and  are given by
\bea
 H^{qq'\bar \ell }_{\rm eff}&=&\frac{4G_F V^*_{qq'}}{\sqrt2}\sum_\ell\Bigg\{\sum_{\chi,\chi'=L,R} \Big[
 \left(\bar q\gamma^\mu[\delta_{\chi'L}\delta_{\chi L}+ C^{V*}_{qq'\ell\chi'\chi}]
 P_{\chi'}q'\right)\
 \left(\bar\nu_\ell\gamma_\mu  P_\chi\ell\right)
 + C^{S*}_{qq'\ell\chi'\chi}\,\left(\bar q
 \gamma^0P_{\chi'}\gamma^0q'\right)\ 
 \left(\bar\nu_\ell  \gamma^0 P_\chi\gamma^0\ell\right)\Big]\no
 &+&\sum_{\chi=L,R} C^{T*}_{qq'\ell\chi\chi}\left(\bar q\,\gamma^0P_\chi\gamma^0\sigma^{\mu\nu}
  q'\right)\ 
 \left(\bar\nu_\ell \gamma^0 P_\chi\gamma^0\sigma_{\mu\nu} \ell\right)\Bigg\}.
 \eea
 Introducing now the charged-conjugated 
 fields\footnote{We have that
 $\Psi^{\cal C}=C\bar\Psi^T$ with $C$ the charge-conjugation matrix that satisfies 
 the relation $C\gamma^{\mu T}C^\dagger=-\gamma^\mu$.} $\nu^{\cal C}_\ell, \ell^{\cal
 C}$, we can rewrite the above terms  as
\bea
 H^{qq'\bar \ell }_{\rm eff}&=&\frac{4G_F V^*_{qq'}}{\sqrt2}\sum_\ell\Bigg\{\sum_{\chi,\chi'=L,R}\Big[
 (\bar q\gamma^\mu[\delta_{\chi'L}\delta_{\chi L}+ C^{V*}_{qq'\ell\chi'\chi}]P_{\chi'}q')\
(\overline{\ell^{\cal C}}[-P_\chi\gamma_\mu]\nu^{\cal C}_\ell)
 + C^{S*}_{qq'\ell\chi'\chi}\,(\bar q
 \gamma^0 P_{\chi'}\gamma^0 q')\ 
 (\overline{\ell^{\cal C}}  \gamma^0 P_\chi\gamma^0\nu^{\cal C}_\ell)\Big]\no
 &+&\sum_{\chi=L,R}  C^{T*}_{qq'\ell\chi\chi}(\bar q\,\gamma^0 P_\chi\gamma^0 \sigma^{\mu\nu}
  q')\ 
 (\overline{\ell^{\cal C}} [-\sigma_{\mu\nu}\gamma^0 P_\chi\gamma^0] 
 \nu^{\cal C}_\ell)\Bigg\}\no
 &=&\frac{2G_F V^*_{qq'}}{\sqrt2}\sum_\ell\sum_{\chi=L,R}\Big[(\bar q[\bar C^V_{qq'\bar\ell\chi}\gamma^\mu+h_\chi 
 \bar C^A_{qq'\bar\ell\chi}
 \gamma^\mu\gamma_5]\,q')\ (\overline{\ell^{\cal C}}\gamma^\mu 
 P_\chi\nu^{\cal C}_\ell)
 +
(\bar q\,[\bar C^S_{qq'\bar\ell\chi} +h_\chi \bar C^P_{qq'\bar\ell\chi}
\gamma_5]\,q') (\overline{\ell^{\cal C}} P_\chi\nu^{\cal C}_\ell)\no
&+&\bar C^T_{qq'\bar\ell\chi}\,(
\bar q\, \sigma^{\mu\nu}[I+h_\chi\gamma_5]\,q')
  (\overline{\ell^{\cal C}} 
\sigma_{\mu\nu}P_\chi\nu^{\cal C}_\ell)\Big],
\label{eq:heffqq'anti}
 \eea 
with
\bea
&&\bar C^V_{qq'\bar\ell L}=-C^{V*}_{qq'\ell R},\quad 
\bar C^A_{qq'\bar\ell L}=C^{A*}_{qq'\ell R},\quad \bar C^V_{qq'\bar\ell R}=-C^{V*}_{qq'\ell L},\quad 
\bar C^A_{qq'\bar\ell R}=C^{A*}_{qq'\ell L}   \no 
&&\bar C^S_{qq'\bar\ell L}= C^{S*}_{qq'\ell R},\quad  \bar C^P_{qq'\bar\ell L}=  C^{P*}_{qq'\ell R},\quad 
\bar C^S_{qq'\bar\ell R}= C^{S*}_{qq'\ell L},\quad  \bar C^P_{qq'\bar\ell R}=  C^{P*}_{qq'\ell L},\no
&& \bar C^T_{qq'\bar\ell L}=- C^{T*}_{qq'\ell R},\quad  \bar C^T_{qq'\bar\ell R}=- C^{T*}_{qq'\ell L},
\label{eq:wqq'anti}
\eea 
and where we have used the relation 
\bea
\overline{\Psi'}\Gamma \Psi=\overline{\Psi^{\cal C}}\,C\Gamma^T C^\dagger\Psi^{\prime{\cal C}},
\eea
 assuming that different fermion fields anti-commute, which implies  
 that $\overline{\Psi'}\Gamma \Psi=- \Psi^T \Gamma^T\, \overline{\Psi'}^T$.
 \footnote{{Note this minus sign arising from the lepton operator was 
 neither taken into account in Eq.~(A.3) of Ref.~\cite{Alvarado:2024lpq}, nor in the Appendix E of Ref.~\cite{Penalva:2021wye}. 
 In this latter case, we directly payed attention to the lepton current constructed out 
 of Dirac spinors and matrices, and not to the lepton operator in terms of fields.   In 
 the present work, this global minus sign has been absorbed into the definition of the WCs in Eq.~(\ref{eq:wqq'anti}) above and is of no relevance to the  calculation of 
 the  hadron tensors which are quadratic in those coefficients.}}

Note the resemblance between Eqs.~(\ref{eq:heffqq'}) and (\ref{eq:heffqq'anti}) and recall that   charge-conjugated fields play for antiparticles the same role as regular fields for particles. Specifically, the fields  $\nu^{\cal C}_\ell$ 
and $\overline{\ell^{\cal C}}$  correspond to the annihilation of $\bar\nu_\ell$ and the creation of $\ell^+$ , respectively, providing $u$-type spinors as wavefunctions in the Dirac space. Thus, the expressions for all nucleon-level reactions $\nu_\ell\,N\to H \ell^-$  (for $d\to u,c$)  and $\bar \nu_\ell\, N\to H' \ell^+$ (for $u\to d,s,b$) for a given spin-parity of the final $H,H'$ hadrons  are formally identical, aside from obvious differences in the CKM  matrix elements,  particle masses, hadronic form factors and WCs.  

General expressions for the hadronic tensors relevant to these reactions have been derived in Ref.~\cite{Penalva:2021wye}, and they are valid for arbitrary spin-parity quantum numbers of the initial and final hadrons. These tensors are constructed from the hadron masses and momenta, and are expressed in terms of structure functions that depend on the form factors, WCs, and masses involved.
Explicit results for the structure functions exist for various transitions, including $0^-\to 0^-,1^-$~\cite{Penalva:2020ftd}, $1/2^+\to 1/2^+$~\cite{Penalva:2020xup} and $1/2^+\to 1/2^-,3/2^-$~\cite{Du:2022ipt}. While the specific form of these structure functions depend on the transition form factors, the overarching formalism remains completely general once the structure functions are determined.

 \section{The $\nu_\tau n\to\Lambda_c\tau^-$ reaction}
In this section we analyze the reaction $\nu_\tau n\to\Lambda_c\tau^-$. As previously mentioned, the expressions derived here are quite general and, beyond this specific case, they serve to illustrate the broader procedure applicable to similar reactions involving other (anti-)neutrino flavors or different charge-exchange quark and antiquark transitions.
The  unpolarized cross section  evaluated in the initial neutron rest-frame (LAB) and neglecting the
  neutrino mass  reads
 \bea
 \sigma=\frac{G_F^2|V_{cd}|^2}{4\pi^2|\vec
 k\,|}\int{d^3p'}\frac{M_{\Lambda_c}}{E'}\int\frac{d^3k'}{k^{\prime0}}
 \delta^{(4)}(k+p-p'-k')\overline\sum|{\cal M}|^2
 \label{eq:sigmafree}
 \eea
 with $k,p,p'$ and $k'$ the four momenta of the  $\nu_\tau$, neutron, 
  $\Lambda_c$ hyperon  and $\tau$ lepton,
 respectively,  and where $\overline \sum$ stands for a sum (average) over the spin or helicities of the final (initial)
 particles. For a fully polarized neutrino beam only a factor 1/2 due to the initial nucleon spin has to be 
 included in the average. This will be the common case since, even in the presence of NP, one expects the neutrino  production to be SM-dominated and, thus, the
 neutrinos will be mainly  produced in a left-handed state which, for high energies compared to its mass\footnote{For this particular  case
  the neutrino threshold energy is given by
 $E_{\nu_\tau}^{\rm th}\approx
 8.3\,$GeV, or in the case of the $\nu_\mu n\to\Lambda_c\mu^-$ reaction 
 $E_{\nu_\mu}^{\rm th}\approx 
 2.6\,$GeV, much larger than the  limits on neutrino masses.},
 implies a well defined negative helicity.  
 If the beam were not fully polarized we would have  to know the neutrino 
 density matrix. For a sufficiently energetic beam we can  identify  
 negative/positive helicities with $L/R$
chiralities, and we would   evaluate
$\overline\Sigma|{\cal M}|^2$ as
\bea
\overline\Sigma|{\cal M}|^2=\frac12\sum_{\chi=L,R}\sum_{\chi'=L,R}
\rho_{\chi'\chi} \sum_{r,r'}\sum_{s'}{\cal M}_{\chi}(p,q,k,k';r,r',s') {\cal M}^*_{\chi'}(p,q,k,k';r,r',s'),
\eea
with  $\rho_{\chi'\chi}$ the neutrino density matrix elements, that satisfy the constraint
$\rho_{LL}+\rho_{RR}=1$, and 
$M_{\chi}(p,q,k,k';r,r',s')$ the transition matrix element that depends 
on the four-momenta
 $k,k',p,q$,  with $q={p-p'}={k'-k}$, the neutrino chirality ($\chi$)
 and the spins $r$ ($r'$) of the initial (final) hadron  and  $s'$
of the final lepton. $M_\chi$
  can be written as
 \be
 {\cal M}_{\chi}(p,q,k,k';r,r',s')=J_{H\chi}^\mu(p,q;r,r') J^L_{\chi\,\mu}(k,k';s')
 +J_{H\chi}(p,q;r,r') J^L_{\chi}(k,k';s')+
 J_{H\chi}^{\mu\nu}(p,q;r,r')J^L_{\chi\,\mu\nu}(k,k';s'),
 \ee
with the
  dimensionful leptonic  currents   given by
 \bea
 J^L_{\chi a}(k,k';s') &=&\frac1{\sqrt2}\bar u_\tau(\vec k\,',s')\Gamma_aP_\chi u_{\nu_\tau}(\vec k,\chi),\nonumber\\
 \hspace{3cm}\Gamma_a&=&\gamma_\mu,I,\sigma_{\mu\nu}
 \eea
and the dimensionless hadronic ones by\footnote{We take the hadron states 
normalization as $\langle \vec{p}\,', r'| \vec{p},
r\rangle= (2\pi)^3(E/M)\delta^3(\vec{p}-\vec{p}\,')\delta_{rr'}$ with $r,r'$  spin indexes.}
\bea
&&J_{H\chi}^a(p,q;r,r')=\langle\Lambda_c;p',r'|\bar c( 0)O^a_{H\chi} d(0)|n;p,r\rangle,\nonumber\\
&&O^a_{H\chi}=C^V_{dc\tau\chi}\gamma^\mu+h_\chi C^A_{dc\tau\chi}\gamma^\mu\gamma_5,\ 
C^S_{dc\tau\chi}+h_\chi C^P_{dc\tau\chi}\gamma_5, \ 
C^T_{dc\tau\chi}\sigma^{\mu\nu}(I+h_\chi\gamma_5),
\eea

From those currents one can build up the leptonic and hadronic
tensors
\bea
L_{\chi ab}(k,k')&=&\sum_{s'}J^L_{\chi a}(k,k';s')J^{L*}_{\chi b}(k,k';s')=\frac14{\rm Tr}[(\slashed{k'}+m_\tau)\Gamma_a(I+h_\chi\gamma_5)\slashed{k}
\gamma^0\Gamma^\dagger_b\gamma^0],\\
W^{ab}_\chi(p,q)&=&\frac12\sum_{r,r'}\langle\Lambda_c;p',r'|\bar c( 0)O^a_{H\chi} d(0)|n;p,r\rangle
\langle\Lambda_c;p',r'|\bar c( 0)O^b_{H\chi} d(0)|n;p,r\rangle^*,\label{eq:ht}
\eea
where   left-right 
interference  terms cancel out exactly if one neglects contributions proportional to the neutrino mass.\footnote{ 
 This is so since, for massless neutrinos, the trace
that leads to the lepton tensor  contains the factor
\bea
P_\chi\slashed{k}\gamma^0P_{\chi'}=P_\chi P_{\chi'}\slashed{k}\gamma^0
=\delta_{\chi\chi'}P_\chi\slashed{k}\gamma^0.
\eea}
 Besides, we have already
included the $1/2$ average factor, due to the initial nucleon spin, in the definition of 
the hadronic tensors.

It terms of the above-defined tensors we have that
\bea
\overline\Sigma|{\cal M}|^2=\sum_{\chi=L,R}\rho_{\chi\chi}\sum_{a,b}W^{ab}_\chi L_{\chi ab}.
\eea
The
 fully left-handed polarized case for neutrino  corresponds to 
 $\rho_{ LL}=1,\,\rho_{RR}=0$, while the opposite will be true for a fully 
 right-handed antineutrino beam, i.e. $\rho_{ LL}=0,\,\rho_{RR}=1$ for this latter case.

The expressions for the different $L_{\chi ab}$ can be obtained from   appendix B of
Ref.~\cite{Penalva:2021wye} by taking twice the unpolarized part of the 
lepton tensors there, that is to say, by summing for both polarizations  of the outgoing tau-lepton.  
As for the hadronic tensors, their expressions  are completely general and they   can be read off 
from  appendix C of Ref.~\cite{Penalva:2021wye}, where they are given for a similar $H_b\to H_c
$ transition.\footnote{In those expressions, $M$ refers to the mass of the 
initial hadron; $M_n$ in the present
 case.   } In fact, we have taken $q=p-p'\,(=k'-k)$,  which is minus the lepton four-momentum tranfer, 
so that one can directly use  appendix C of 
Ref.~\cite{Penalva:2021wye} for the hadronic part of the amplitude. The hadronic
tensors
 depend on the
$\widetilde W_\chi$ structure functions that are built out of the two hadron 
masses,  the WCs
and the form factors that parametrize the hadronic matrix elements of the different quark-current 
operators. For this particular case we  are dealing with
a $1/2^+\to 1/2^+$ transition for which we take the form factor parametrization
\begin{eqnarray}
 \langle\Lambda_c;{p}^{\,\prime},r'| \bar c(0) d(0)|
 n;{p};r\rangle &=& F_S\,\bar{u}_{\Lambda_c}
 (\vec{p}^{\,\prime},r'\,)
 u_{n}(\vec{p},r)\no
 \langle\Lambda_c;{p}^{\,\prime},r'| \bar c(0)\gamma_5 d(0)|
 n;{p};r\rangle &=&  F_P\,\bar{u}_{\Lambda_c}
 (\vec{p}^{\,\prime},r'\,)\gamma_5 
 u_{n}(\vec{p},r), \nonumber \\
\langle\Lambda_c;{p}^{\,\prime},r'| \bar c(0) \gamma^\alpha d(0) 
|n;\vec{p},r\rangle
&=&
\bar{u}_{\Lambda_c}(\vec{p}^{\,\prime},r')\Big( \gamma^\alpha 
F_1  + \frac{p^{\alpha}}{M_{n}} 
F_2 + \frac{p^{\prime\alpha}}{M_{\Lambda_c}}F_3 \Big) 
 u_{n}(\vec{p},r) \nonumber \\
\langle\Lambda_c;{p}^{\,\prime},r'| \bar c(0)
\gamma^\alpha\gamma_5d(0) 
|n;\vec{p},r\rangle
&=&
\bar{u}_{\Lambda_c}(\vec{p}^{\,\prime},r')\Big( \gamma^\alpha 
\gamma_5G_1  + \frac{p^{\alpha}}{M_{n}} 
\gamma_5G_2 + \frac{p^{\prime\alpha}}{M_{\Lambda_c}}\gamma_5G_3 \Big) 
 u_{n}(\vec{p},r) \nonumber \\
\langle\Lambda_c;{p}^{\,\prime},r'|  \bar c(0)\sigma^{\alpha\beta}d(0)   
|n;\vec{p},r\rangle &=&
 \bar{u}_{\Lambda_c}(\vec{p}^{\,\prime},r')
 \Big\{i\frac{T_1}{M^2_{n}}
 \left(p^\alpha p^{\prime \beta}-p^\beta p^{\prime \alpha}\right) 
 + i\frac{T_2}{M_{n}}\left(\gamma^\alpha p^\beta
  -\gamma^\beta p^\alpha  \right) \nonumber \\
&&\hspace{2cm}+ i\frac{T_3}{M_{n}}\left(\gamma^\alpha p^{\prime \beta}
-\gamma^\beta p^{\prime \alpha}\right) 
+ T_4 \sigma^{\alpha\beta} \Big\} u_{n}(\vec{p},r),
\label{eq.FactoresForma}
\end{eqnarray}

The corresponding expressions for the $\widetilde W_\chi$ structure
functions in terms of the WCs and the above form factors  are collected in 
appendix E of Ref.~\cite{Penalva:2020xup}, where one has to change 
$M_{\Lambda_b}$ and $C_{A,V,S,P,T}$ there by $M_n$ and $C^{A,V,S,P,T}_{dc\tau\chi}$ 
respectively. As  mentioned above,  the relations between the $\widetilde W_\chi$ structure
functions and the corresponding form factors are also available  for
 $0^-\to 0^-,1^-$~\cite{Penalva:2020ftd},  and 
 $1/2^+\to 1/2^-,3/2^-$~\cite{Du:2022ipt} transitions.

The  form factors in Eq.~(\ref{eq.FactoresForma}) can be related to the $\Lambda_c\to n$ helicity form 
factors of Ref.~\cite{Meinel:2017ggx} where a LQCD determination of the
 vector, axial and tensor form factors has been carried out for that decay. 
 The relation between both form factor sets can be found here in Appendix \ref{app:ff}. 
 The form factors in  Ref.~\cite{Meinel:2017ggx}, have to 
be analytically continued
 from  the $q^2>0$ region of the $\Lambda_c\to n$ decay to 
the $q^2<0$ one of the $\nu_\tau n\to \Lambda_c\tau^-$ reaction. This
extrapolation outside  the fitted region is
clearly a source of uncertainty since the two $q^2$ regions are far apart.\footnote{ 
Even at the
threshold for the $\nu_\tau n\to \Lambda_c\tau$ reaction one already has 
$q^2\approx-3.68$\,GeV$^2$, while in
Ref.~\cite{Meinel:2017ggx} the form factors are evaluated on the lattice for 
$q^2\gtrsim -0.36$\,GeV$^2$.} This effect can be seen for instance in Fig. 16 of
Ref.~\cite{Kong:2023kkd}.

Similarly to what we obtained for a $H\to H'\tau^-\nu_\tau$ decay (see
Eq.~(14) of Ref.~\cite{Penalva:2020xup}), we now
find that $\overline\Sigma|{\cal M}|^2$ can be written as\footnote{The sign changes
 follow from crossing symmetry, taking into account that between the decay and the 
scattering reaction, one fermion external leg (neutrino of four momentum $k$) 
 is interchanged between  final and initial configurations. }
\bea
\overline\Sigma|{\cal M}|^2=\frac{M^2_n}2\Big[-{\cal A}(\omega)+{\cal B}(\omega)
\frac{p\cdot k}{M^2_n}-{\cal C}(\omega)
\frac{(p\cdot k)^2}{M^4_n}\Big], \label{eq:ampli-hadron}
\eea
with $\omega= (p\cdot p')/(M_n M_{\Lambda_c})=
(M^2_{\Lambda_c}+M^2_n-q^2)/(2M_{\Lambda_c}M_n)$ the product of the two
hadron four-velocities.    The expressions for 
the ${\cal A}(\omega), {\cal B}(\omega)$ and ${\cal C}(\omega)$ functions  in 
terms of the $\widetilde W_\chi$ structure functions
are given  in appendix D of Ref.~\cite{Penalva:2021wye}, where $M, M_\omega$ and $m_\ell$ there
correspond to $M_n, (M_n-\omega M_{\Lambda_c})$ and $m_\tau$ respectively.
Moreover, it is necessary to include  a factor $\rho_{\chi\chi}$ in the corresponding sums over neutrino chiralities present in the definition of ${\cal A}(\omega), {\cal B}(\omega)$ and ${\cal C}(\omega)$. This factor was not required in Ref.~\cite{Penalva:2021wye}, where the neutrino was an outgoing particle. In the present case, however, the neutrino is an incoming one, and that is why the $\rho_{\chi\chi}$ factor should be included. For a purely left-handed neutrino beam, one has $\rho_{{LL}}=1,\,\rho_{{RR}}=0$, so only the left-handed chirality terms contribute to the ${\cal A}(\omega), {\cal B}(\omega)$ and ${\cal C}(\omega)$ functions.

After integration of the delta function, we get
\bea
\sigma=\frac{G^2_F|V_{cd}|^2M^2_nM^2_{\Lambda_c}}{4\pi|\vec k\,|^2}
\int_{\omega_{-}}^{\omega_{+}}d\omega\,\Big[-{\cal A}(\omega)+{\cal B}(\omega)
\frac{|\vec k\,|}{M_n}-{\cal C}(\omega)
\frac{|\vec k\,|^2}{M^2_n}\Big],
\eea
with $\omega_\pm
={E'_{\pm}}/{M_{\Lambda_c}}$ and
\bea
E'_{\pm}=\frac1{2SM_{\Lambda_c}}\left[(|\vec k\,|+M_n)(S+M^2_{\Lambda_c}-m^2_\tau)\pm|\,\vec
k\,|\, \lambda^{1/2}(S,M^2_{\Lambda_c},m^2_\tau)\right] \label{eq:eprimemaxmin}
\eea
the maximum and minimum energies allowed  for the final $\Lambda_c$ hadron in the LAB
reference frame, corresponding, respectively, to the $\Lambda_c$ being emitted in the
forward/backward direction in the  neutrino-nucleon center of mass  frame.
 In addition, $S=M_n^2+2M_n|\vec k\,|$ is the total center of mass energy squared
and $\lambda(x,y,z)=z^2+y^2+z^2-2xy-2xz-2yz$,  the K\"allen lambda
function.

The $d\sigma/dQ^2$ differential cross section is  given by (with $Q^2=-q^2$)
\bea
\frac{d\sigma}{dQ^2}=\frac1{2M_{\Lambda_c}M_n}\frac{d\sigma}{d\omega}=
\frac{G^2_F|V_{cd}|^2M_nM_{\Lambda_c}}{8\pi|\vec k\,|^2}
\Big[-{\cal A}(\omega)+{\cal B}(\omega)
\frac{|\vec k\,|}{M_n}-{\cal C}(\omega)
\frac{|\vec k\,|^2}{M^2_n}\Big].
\eea

\begin{figure}[htb]
\begin{center}
\includegraphics[height=.35\textwidth]{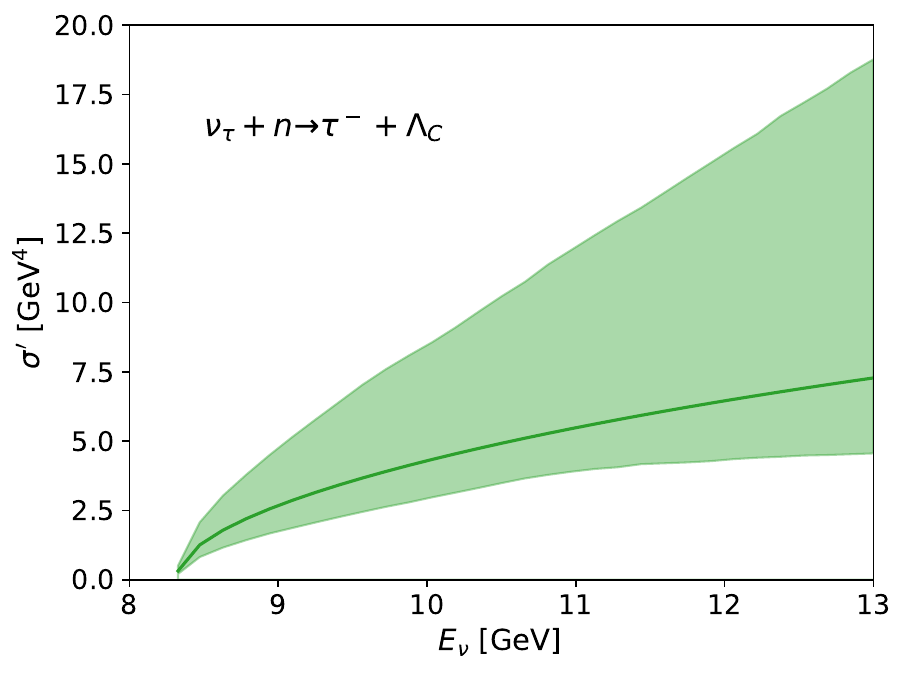}\hspace{.5cm} 
\includegraphics[height=.35\textwidth]{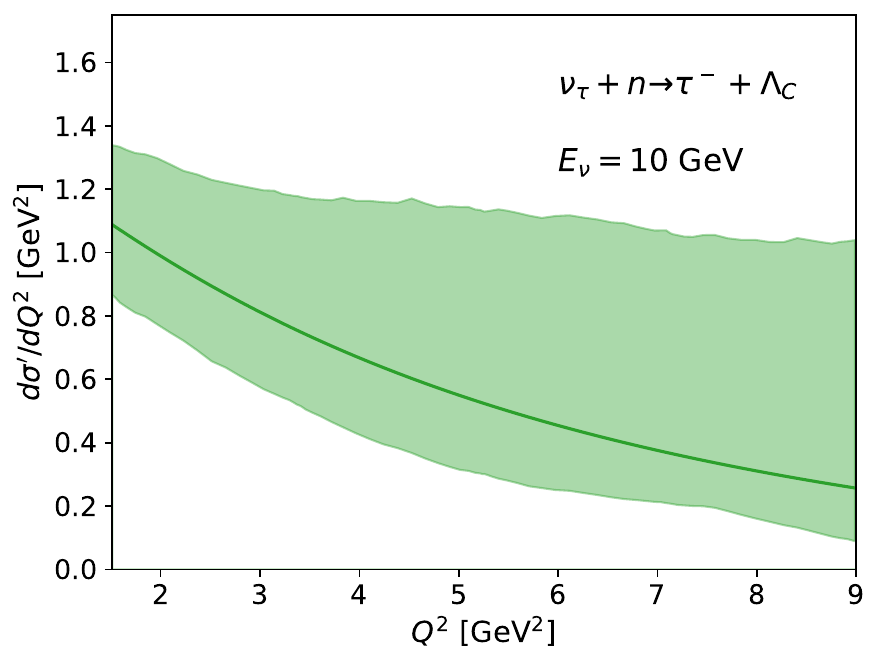}
\caption{ The $\nu_\tau n\to \Lambda_c \tau^-$ reaction evaluated within the SM  for a fully left-handed polarized  $\nu_\tau$ beam. Left panel: $\sigma'=(8\pi M_n^2\sigma)/(G_F|V_{cd}|)^2$ cross section as a function of the neutrino energy. Right panel:
$d\sigma'/dQ^2$ differential distribution  for $E_{\nu_\tau}=10\,$GeV.
}  
\label{fig:sigmap}
\end{center}
\end{figure}

In Fig.~\ref{fig:sigmap}, we show the results for the scaled cross-section $\sigma'=(8\pi M_n^2\sigma)/(G_F|V_{cd}|)^2$ and $d\sigma'/dQ^2$, evaluated within the SM considering the neutrino beam is fully (left-handed) polarized. Focusing on the central values, our results are a factor of two smaller than those reported in Ref.~\cite{Yan:2024bce}. We suspect this discrepancy may arise from an error in the analysis presented in that reference. The green band represents the total uncertainty associated with the determination of the form factors. As evident from the figure, this uncertainty significantly hinders the possibility of making precise theoretical predictions for the reaction. 
Even under the idealized assumption of perfectly known form factors, one must still contend with the fact that the process occurs within nuclear targets, where nuclear effects introduce additional theoretical uncertainties. These are examined in detail in the following section.

\section{The  $\nu_\tau n\to \Lambda_c\tau^- $ QE reaction in nuclei}
\label{sec:nucleus}
\subsection{General considerations}

To address the nuclear effects for the case of the $\Lambda_C$ QE production in nuclei, we assume the impulse approximation, i.e. the electroweak probe scatters on individual nucleons which form the nucleus, and the total cross section is an incoherent sum of these single-nucleon contributions.
The sum can be performed either at the level of cross sections or at the level of the hadronic tensors. These two descriptions would be equivalent if all the particles in the reaction were on-shell and the transferred energy-momentum to the nuclear system was fully transferred to an interacting nucleon.
A well established approach to address the nuclear effects of the ground-state nucleus is the use of a hole spectral function (SF) $S_h^{(n/p)}( E,\vec p\,)$, which yields the probability density of finding a neutron/proton in the target nucleus with a given momentum $\vec p$ and energy $E$. One expects the SF to be
rotational invariant for a nucleus at rest, i.e. $S_h^{(n)}(E,\vec p\,)=S_h^{(n)}(E,|\vec
p\,|)$. The final state interactions (FSI) which affect the outgoing hadrons, can be currently described only in  a very simplified way for the $\Lambda_c$ particle.

 We start from the cross section of Eq.\eqref{eq:sigmafree} for the scattering  of a neutrino of energy $ |\vec{k}|$ off a free nucleon at rest
\bea
 \sigma=\frac{G_F^2|V_{cd}|^2}{4\pi^2|\vec
 k\,|}\int{d^3p'}\frac{M_{\Lambda_c}}{E'}\int\frac{d^3k'}{k^{\prime0}}
 \delta^{(4)}(k+p-p'-k')\overline\sum|{\cal M}|^2\,.
\eea
Now the sum over all neutrons in the nucleus $A_Z$ corresponds to the dimensionless integral over the spectral function which is normalized as
\bea
\int\frac{d^3p}{(2\pi)^3}\int d E\, S_h^{(n)}( E,\vec p\,)=A-Z
\eea
Neglecting the effects of the $\Lambda_c$ interactions with the nuclear environment where it is produced, we have
\bea
\sigma_{A_Z}
&=&\frac{G_F^2|V_{cd}|^2}{4\pi^2|\vec
 k\,|}\int\frac{d^3p}{(2\pi)^3}\int dE\frac{M_n}{\sqrt{M_n^2+\vec{p}^{\,2}}}S_h^{(n)}(E,|\vec p\,|)\int{d^3p'}\frac{M_{\Lambda_c}}{E'}\int\frac{d^3k'}{k^{\prime0}}
 \delta^{(4)}(k+p-p'-k')\,\overline\Sigma|{\cal M}|^2 \no
&=&\frac{G_F^2|V_{cd}|^2}{4\pi^2|\vec
 k\,|}\int\frac{d^3k'}{k^{\prime0}}
 F^{A_Z}_{\nu_\tau \Lambda_c}(q),
 \label{eq:sigaz}
\eea
where we have introduced the  squared modulus of the spin-averaged nuclear amplitude $F^{A_Z}_{\nu_\tau \Lambda_c}(q)$  (with dimension of energy in natural units),
\be 
F^{A_Z}_{\nu_\tau \Lambda_c}(q) =\int\frac{d^3p}{(2\pi)^3}\int dE\,S_h^{(n)}(E,|\vec p\,|)\frac{M_n}{\sqrt{M_n^2+\vec{p}^{\,2}}}\frac{M_{\Lambda_c}}{E'}\delta(E-E'-q^0)\,\overline\Sigma|{\cal M}|^2 .
\label{eq:faz-def}
\ee
The ${M_n}/{\sqrt{M_n^2+\vec{p}^{\,2}}}$ factor in the above equations accounts for the covariant normalization of the Dirac-spinor of the initial nucleon when is not at rest, while the on-shell energies of the initial nucleon and the final hadron are $\sqrt{M_n^2+\vec p\,^2}$ and $E'=\sqrt{M_{\Lambda_c}^2+(\vec p-\vec q\,)^2}$, respectively.

Next, we consider two different procedures to compute the nuclear response function  $F^{A_Z}_{\nu_\tau \Lambda_c}(q)$, whose differences could provide an estimate of the theoretical uncertainties associated with our predictions for the $\Lambda_c$ production in nuclei. 
\begin{itemize}
    \item {\it Nuclear corrections by convoluting the single-nucleon cross section:} In this case, we approximate   
    \be 
F^{A_Z}_{\nu_\tau \Lambda_c}(q) =\int\frac{d^3p}{(2\pi)^3}\int dE\,S_h^{(n)}(E,|\vec p\,|)\frac{M_n}{\sqrt{M_n^2+\vec{p}^{\,2}}}\frac{M_{\Lambda_c}}{E'}\delta(E-E'-q^0)\frac{M_n^2}2\Big(-{\cal A}(\omega)+{\cal B}(\omega)\frac{p\cdot k}{M_n^2}-{\cal C}(\omega)\frac{(p\cdot k)^2}{M_n^2}\Big).
\label{eq:faz1}
\ee
Note that in the computation of $F^{A_Z}_{\nu_\tau \Lambda_c}(q)$, in this scenario we first pre-calculate $\overline{\sum}|{\cal M}|^2$ at the hadron level, see Eq.~\eqref{eq:ampli-hadron}, assuming that $p^2=M_n^2$ and $p'^2=M_{\Lambda_c}^2$. Effectively, the only explicit dependence on $p$ appears in the four-scalar product $(p\cdot k)$. The integral over the hole SF affects the energy conservation, giving rise to the  allowed $(q^0,|\vec{q}\,|)$ domain of this QE reaction. The inclusion of the SF also leads to a more realistic evaluation in a nuclear medium of the $(p \cdot k)$ and $(p\cdot k)^2$ terms that appear in Eq.~\eqref{eq:ampli-hadron} where we take $p^0=E$ off the mass-shell, and thus  independent of $\vec p$. A detailed way of evaluating Eq.~\eqref{eq:faz1} is  explained in Appendix~\ref{app:a012b01c0d0}. 

    \item {\it Nuclear corrections by convoluting the single-nucleon hadron tensor:} In this case the squared modulus of the spin-averaged nuclear amplitude is calculated as
\bea
F^{A_Z}_{\nu_\tau \Lambda_c}(q)=\sum_{\chi=L,R}\rho_{\chi\chi}\sum_{a,b}W_{A_Z; \chi}^{ab}(q)\,L_{\chi\,ab}(k,q).
\label{eq:faz2}
\eea
with the nuclear tensor\footnote{This definition  is in accordance with the one in  Refs.~\cite{Nieves:2004wx,Valverde:2006yi,Sobczyk:2019uej,Hernandez:2022nmp}, 
except that in these latter cases the corresponding  $W^{ab}_{A_Z\chi}$ tensors 
include the CKM matrix element squared. Note also that in 
Refs.~\cite{Nieves:2004wx,Valverde:2006yi,Sobczyk:2019uej,Hernandez:2022nmp} only SM tensors are considered,
i.e. the ones generated from the vector-minus-axial current alone.},  constructed out of the hadron-level tensor $W^{ab}_\chi(p,q)$ of Eq.~\eqref{eq:ht},
\bea
W_{A_Z; \chi}^{ab}(q)=\int\frac{d^3p}{(2\pi)^3}\int dE \, S_h(E,|\vec p\,|)
\frac{M_n}{\sqrt{M_n^2+\vec{p}^{\,2}}}\frac{M_{\Lambda_c}}{E'}\,
 \delta( E-E'-q^0)W^{ab}_\chi(p,\tilde{q})\Big|_{\tilde q = \left(q_0+E-\sqrt{M_n^2+\vec{p}^{\,2}},\, \vec{q}\,\right)}\,
 \label{eq:nucten_SF}
\eea
which assumes that both the initial nucleon and the final $\Lambda_c$ are on-shell in the matrix element at the single-nucleon level. This is achieved by modifying the energy transferred to the nucleon, so that the energy conservation reads
\bea
\delta(E-E'-q^0) \equiv \delta(\sqrt{M_n^2+\vec{p}^{\,2}}-E'-\tilde q^0) 
\eea

\end{itemize}

We note that in both descriptions above, the nuclear response  $F^{A_Z}_{\nu_\tau \Lambda_c}(q)$ is rotational invariant. Therefore, it can only depend on $|\vec k\,|,\,|\vec k\,'|$ and $\vec k\cdot\vec k\,'=|\vec k\,|
 |\vec k\,'|\cos\theta_\tau$, and one can  trivially integrate in $\varphi_\tau$
 in Eq.~\eqref{eq:sigaz} to get
\bea
\sigma_{A_Z}
=\frac{G_F^2|V_{cd}|^2}{2\pi|\vec
 k\,|}\int_{m_\tau}^{|\vec k\,|+\Delta M-M_{\Lambda_c}} |\vec k\,'| 
 \,dk^{\prime0}\int_{-1}^{+1} d\cos\theta_\tau\  F^{A_Z}_{\nu_\tau \Lambda_c}(q),
\label{eq:secQEnucleo} 
\eea
 with $\Delta M$ the difference between the mass $M_{A_Z}$ of 
the initial nucleus and the energy of the lowest state of the final $(A-1)_Z$ 
nuclear system. In Eq.~\eqref{eq:secQEnucleo}, for the tau kinematics, we have considered the largest possible  phase space  admitting that the final nuclear system can accommodate any 
 momentum transfer. In a real calculation, the  nuclear amplitude 
 $F^{A_Z}_{\nu_\tau \Lambda_c}(q)$ will limit the actual phase space that 
 can be accessed. In the case of SM currents producing a proton instead of 
 a $\Lambda_c^+$, Eq.~\eqref{eq:secQEnucleo} reduces to Eq.~(2) of
  Ref.~\cite{Hernandez:2022nmp} taking into account that the 
  $F_{\nu_\tau}$ nuclear QE amplitude introduced there corresponds to 
  $|V_{ud}|^2F^{A_Z}_{\nu_\tau}(q)/(2|\vec{k}\,|M_{A_Z})$, within the 
  normalizations adopted in this work. 
  Note also that the $F_{\nu_\tau}$ nuclear 
  QE amplitude in Ref.~\cite{Hernandez:2022nmp} includes  Pauli blocking of the
  final nucleon.
 
One can now evaluate the $d\sigma_{A_Z}/dQ^2$ differential cross section as
\bea
\frac{d\sigma_{A_Z}}{dQ^2}
&=&\frac{G_F^2|V_{cd}|^2}{2\pi|\vec
 k\,|}\int_{m_\tau}^{|\vec k\,|+\Delta M-M_{\Lambda_c}} |\vec k\,'| 
\, dk^{\prime0}\int_{-1}^{+1} d\cos\theta_\tau\ \delta(Q^2+m^2_\tau-2|\vec k\,|
 (k^{\prime0}-|\vec k\,'|\cos\theta_\tau))\,F^{A_Z}_{\nu_\tau \Lambda_c}(q)\no
 &=&\frac{G_F^2|V_{cd}|^2}{4\pi|\vec
 k\,|^2}\int_{m_\tau}^{|\vec k\,|+\Delta M-M_{\Lambda_c}} 
 dk^{\prime0} H(1-|\cos\theta^0_\tau|)\,F^{A_Z}_{\nu_\tau
 \Lambda_c}(q,k)|_{\cos\theta_\tau=\cos\theta^0_\tau},
\label{eq:dsecdq^2QE} 
\eea
with
\bea
\cos\theta^0_\tau=\frac{Q^2+m^2_\tau-2|\vec k\,|k^{\prime0}}{2|\vec k\,||\vec
k\,'|}.
\eea
The $H(1-|\cos\theta^0_\tau|)$ Heaviside function imposes the constraint 
\bea
k^{\prime0}\ge\frac{(Q^2+m^2_\tau)^2+4 m^2_\tau|\vec k\,|^2}{4|\vec
k\,|(m^2_\tau+Q^2)}.
\eea

\subsection{Spectral functions}
\label{sec:SF}
In our calculations, we employ two nuclear hole SFs. We will start with the simple description of the nucleus through the local Fermi gas approximation (LFG) and later move to a more realistic SF.
\subsubsection{LFG approximation}

Within the LFG model the neutron-hole structure function is given by~\cite{Sobczyk:2019uej}
\bea
S_{h}^{\rm LFG}(E,|\vec p\,|)&=&2\int d^3r\,H\left[k_F(r)-|\vec p\,|\right]\delta\left(E-
\sqrt{M_n^2+\vec{p}^{\,2}}\right) 
\label{eq:ShLFG}
\eea
with $k_F(r)=(3\pi^2\rho_n(r))^{1/3}$  and $\rho_n(r)$ the local  neutron Fermi momentum and density distribution, respectively. For an isospin symmetric nucleus, the latter is directly related to the nucleon density, $\rho_n(r) = \rho(r)/2$.  Strictly speaking, in a LFG, the nucleons are moving in an attractive central 
potential $-k_F^2(r)/2M$, with $M$ an averaged nucleon mass, which is not being reflected 
in the energy conservation  delta function of Eq.~\eqref{eq:ShLFG}. However, this 
is a sufficiently accurate approximation as long as it is reasonable to assume that the final particle feels the same attractive potential, which will cancel in the delta of energy conservation
$\delta(E-E'-q^0)$.  In this approximation, the hadrons are treated on-shell (see  Eq.~\eqref{eq:ShLFG}), and the two approaches discussed previously in Eqs.~\eqref{eq:faz1} and \eqref{eq:faz2} to compute the nuclear response $F^{A_Z}_{\nu_\tau \Lambda_c}(q,k)$ are equivalent.

\subsubsection{Realistic spectral function}
\label{sec:realSF}
Among several theoretical approaches proposed to obtain the hole SF~\cite{Benhar:1994hw,Rocco:2018vbf,Nieves:2017lij}, here we will follow  the state-of-the-art ab initio calculation from Ref.~\cite{Sobczyk:2023mey} which employs realistic nuclear many-body wavefunctions, obtained within the coupled cluster theory~\cite{Hagen:2013nca}, and a nuclear Hamiltonian derived in chiral effective field theory. 
This approach has been first benchmarked for $^4$He~\cite{Sobczyk:2022ezo} and  later extended to $^{16}$O. In both cases, a successful comparison with experimental $(e,e')$ data  has been obtained.
The hole SF is defined in the second quantization as
\begin{equation}
    S(E,p) = \sum_{\alpha,\beta} \langle p|\alpha\rangle ^\dagger \langle p|\beta\rangle \sum_{\Psi_{A-1}} \langle \Psi_0|a_\alpha^\dagger|  \Psi_{A-1}\rangle \langle\Psi_{A-1}| a_\beta |\Psi_0\rangle\, \delta(E_0-E-E_{A-1})\ ,
    \label{eq:SF_CC}
\end{equation}
where $\Psi_0$ is a correlated many-body $A$-particle ground-state wavefunction. The sum is made over $\alpha$, $\beta$, the quantum numbers of single-particle states which form a basis of the calculation; $a_\alpha^\dagger$, $a_\alpha$ being the corresponding creation/annihilation operators. The sum  is dominated by the diagonal term which, together with the integral over all $\Psi_{A-1}$ intermediate states, gives rise to the nuclear shell structure. Additional fragmented strength at higher energies comes from the correlations in the wavefunction.
In addition, $\langle p|\alpha \rangle$ denotes single-particle wave functions in  momentum space.
The calculation is performed within the coupled cluster theory. In this approach, the many-body Schr\"odinger equation is solved following an ansatz for the ground state wavefunction $|\Psi_0\rangle = e^T |0\rangle $. Here, we start from a reference state $|0\rangle$ (typically corresponding to a Hartree-Fock solution), and include further correlations by means of the $e^T$ operator, where  $T$  is a sum of excitations
\begin{equation}
    T=\sum_{a,i} t^a_i\, a_a^\dagger a_i + \frac{1}{4} \sum_{a,b,i,j} t^{ab}_{ij}\, a_a^\dagger a_b^\dagger a_i a_j + \cdots
\end{equation}
Indices $a,b$ correspond to particles states, while $i,j$ to hole states. The sum is expanded in particle-hole, two-particle two-hole, etc., excitations and is truncated at a certain level. The SFs which we use here are obtained within a scheme that considers coupled-cluster single and double excitations. The $t$ amplitudes are determined by solving a set of coupled non-linear equations. 
Since the $e^T$ operator is non-Hermitian, the coupled cluster framework requires an additional de-excitation $\Lambda$ operator when defining the left Hamiltonian eigenstates $\langle \Psi_0 | = \langle 0|(1+\Lambda)e^{-T}$. 
The excited spectrum $\Psi_{A-1}$ in Eq.~\eqref{eq:SF_CC} is reconstructed via the Gaussian integral transform expanded on a basis of Chebyshev polynomials~\cite{Sobczyk:2021ejs} using the equation-of-motion technique~\cite{Stanton:1993vcu}. 
The nuclear dynamics is described by a chiral  NNLO$_\text{sat}$ Hamiltonian~\cite{Ekstrom:2015rta}, which contains nucleon-nucleon and three-nucleon forces at the next-to-next-to leading order of chiral effective field theory. The low energy constants are fitted to nucleon-nucleon scattering data as well as to some properties of light and medium-mass nuclei. This theoretical setup allows to perform robust uncertainty studies both in terms of nuclear interaction, as well as the many-body truncations.

\subsection{FSI}
Due to the lack of information on the $\Lambda_c$ properties inside of a nuclear medium, we account for FSI effects by including a  binding energy in its production. This leads to a modification of the energy conservation delta function in Eqs.~\eqref{eq:faz1} and~\eqref{eq:nucten_SF} where we add a constant binding $B_{\Lambda_c}$
   \bea
   \delta(E-E'-q^0)\to\delta(E-(E'-B_{\Lambda_c})-q^0)\,.
   \label{eq:deltamod}
   \eea
and we will estimate the sensitivity of our results by varying the magnitude of the binding energy.

Within the LFG approximation, since we ignore the  $-k_F^2(r)/2M$ potential felt by the nucleon and we use  the $\delta[E'-(E-q^0)]$ energy-conserving delta function for the hadron vertex \cite{Nieves:2017lij,Sobczyk:2019uej}, we are implicitly assuming that the produced  $\Lambda_c^+$ hyperon feels the same local binding potential felt by the nucleon. 

\section{Results}
\label{sec:results}


\subsection{SM results}

In the following, we will focus on the $\nu_\ell +^{16}\rm{O}\to \ell^-+\Lambda_c+X$  reaction, setting WCs in Eq.~\eqref{eq:heffqq'tot} to zero and considering fully left-handed polarized neutrinos.
\begin{figure}[htb]
\begin{center}
\includegraphics[width=0.45\textwidth]{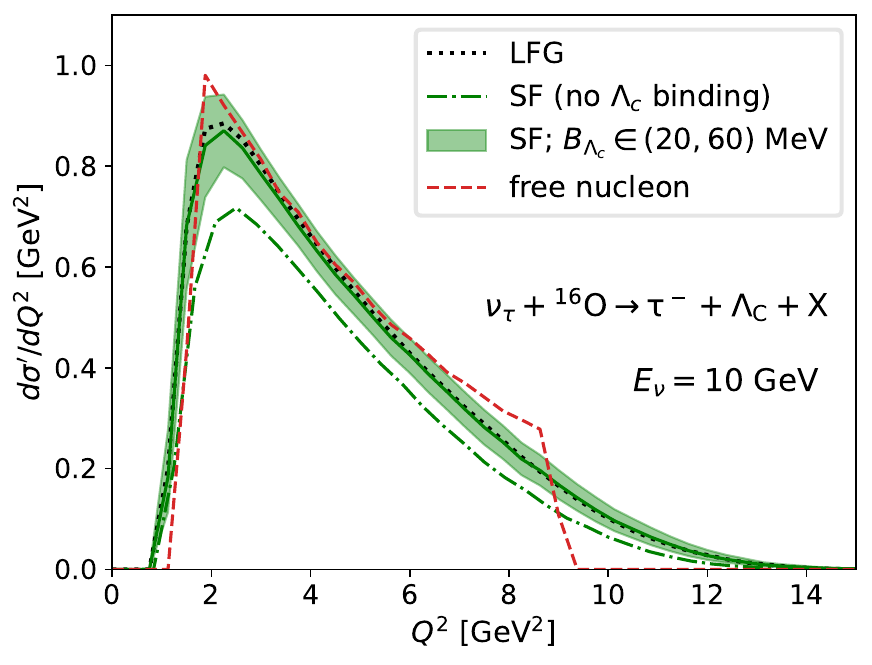}
\includegraphics[width=0.45\textwidth]{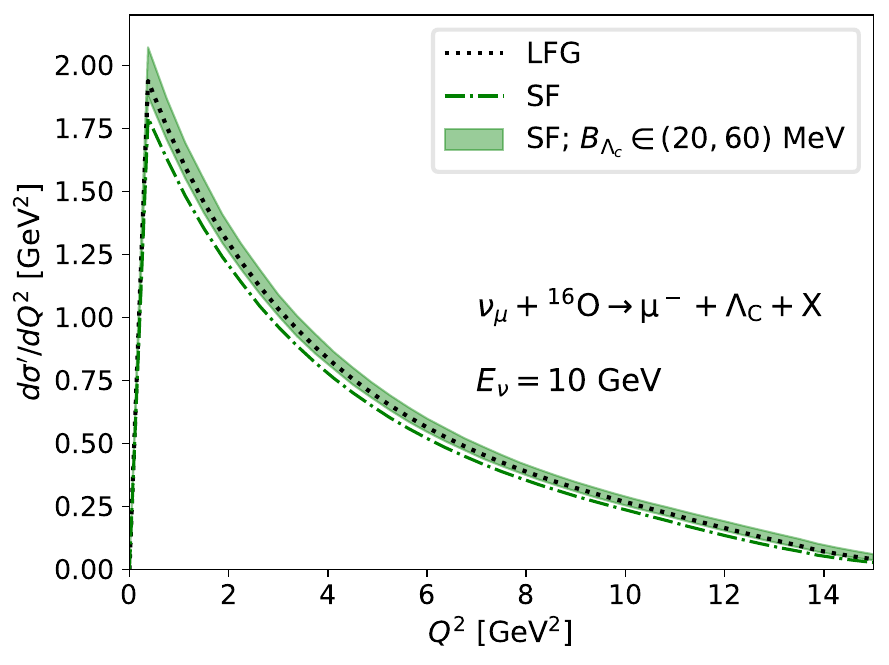}
\caption{ $(d\sigma'/dQ^2)$ cross section per number of neutrons for  the 
QE  reaction $\nu_\ell+ ^{16}\rm{O}\to \ell^-+\Lambda_c+X$ induced by both tau (left) and muon (right) 
neutrinos  evaluated within the SM and for $E_\nu=10\,$GeV.  }  
\label{fig:res_dq2_10GeV}
\end{center}
\end{figure}
In Fig.~\ref{fig:res_dq2_10GeV}, we  show the scaled differential cross section 
$d\sigma'/dQ^2$ normalized per number of neutrons and such that  $\sigma'=(8\pi M_n^2\sigma)/(G_F|V_{cd}|)^2$. We study the reaction induced by both tau (left) and muon (right) neutrinos of 10\, GeV energy. Looking at the $\nu_\tau$ induced reaction we see the appearance of tails for low and high $Q^2$ (below $1$ and above $9$ GeV$^2$, respectively) in the nuclear reaction when compared to the free neutron case (shown as a red dashed curve in the figure). The computation using a realistic hole SF leads to a quenching of the differential cross section with  respect to the  LFG distribution. The agreement between the two calculations clearly improves if we include binding to the $\Lambda_c$, and it is remarkably good for $B_{\Lambda_c}=40$ MeV (solid green line). This is somehow expected since, as mentioned, the LFG calculation implicitly assumes a local binding for the $\Lambda_c$ which  equals  the one felt by the nucleon. In the figure we show a theoretical uncertainty band associated to the unknown binding of the $\Lambda_c$ in $^{16}$O. Similar results are obtained for the $\nu_\mu$ induced reaction. Here, however, the uncertainty and the difference between various predictions is less significant.

\begin{figure}[htb]
\begin{center}
\includegraphics[width=0.45\textwidth]{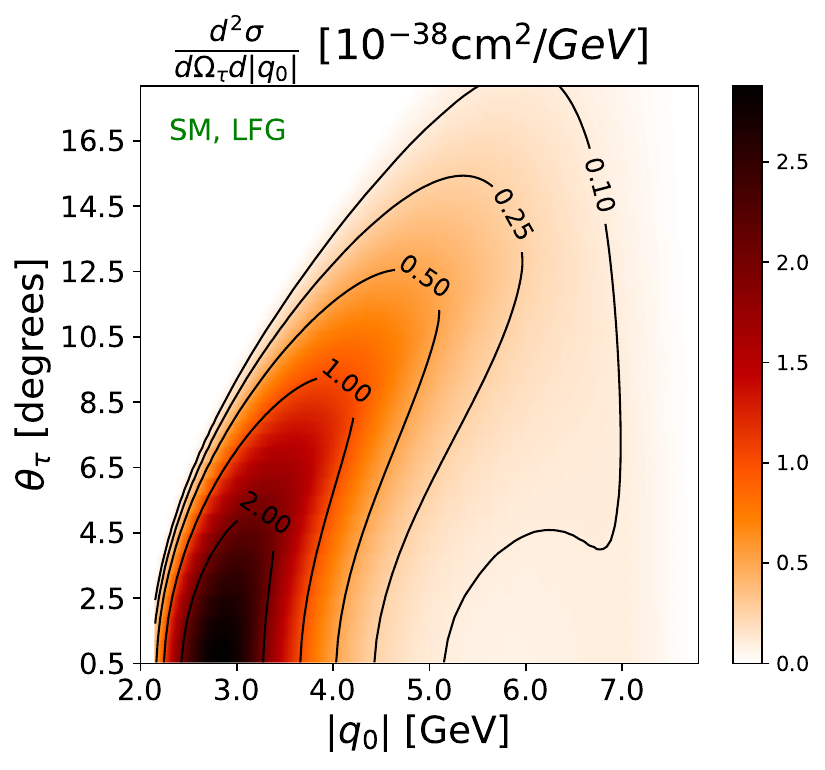}
\includegraphics[width=0.45\textwidth]{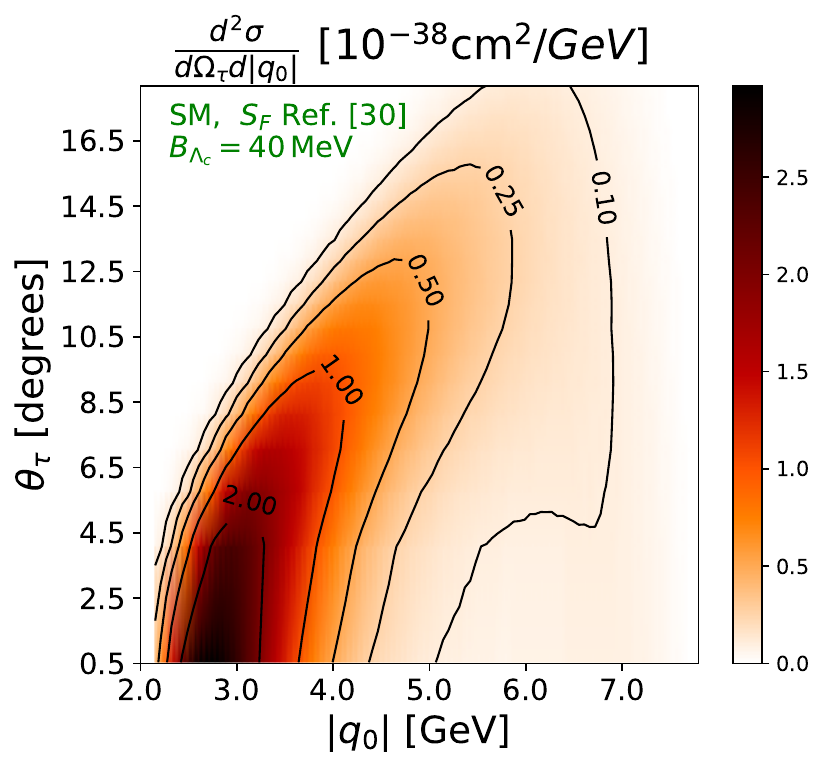}
\caption{ Two dimensional $d^2\sigma/(d\Omega_\tau d|q_0|)$  differential cross section  for  the  QE  reaction $\nu_\tau+ ^{16}\rm{O}\to \tau^-+\Lambda_c+X$ evaluated within the SM and for $E_\nu=10\,$GeV. The left panel has been obtained using the LFG hole SF, while for the right one, we have used the realistic hole SF described in Susbsec.~\ref{sec:realSF}, with $B_{\Lambda_c}= 40$ MeV and the nuclear response $F^{A_Z}_{\nu_\tau \Lambda_c}(q)$  evaluated as indicated in Eq.~\eqref{eq:faz2}. }  
\label{fig:res_2D_10GeV}
\end{center}
\end{figure}

In Fig.~\ref{fig:res_2D_10GeV} we show the $\nu_\tau+ ^{16}\rm{O}\to \tau^-+\Lambda_c+X$ two-dimensional $d^2\sigma/(d\Omega_\tau d|q_0|)$  differential cross section for $E_\nu=10\,$GeV, and evaluated using both the LFG (left) and the ab initio realistic (right)  hole SFs, the latter using  $B_{\Lambda_c}=40$ MeV. We see that the reaction is very much forward peaked ($\theta_\tau \lessapprox 15^{\rm o}$) and that for this neutrino energy, the  signals are obtained for energy-transfers below 4-5 GeV. Moreover, we also observe that the LFG approach provides a quite good description of the results obtained with the realistic SF if we assume $B_{\Lambda_c}=40$ MeV. 

\begin{figure}[htb]
\begin{center}
\makebox[0pt]{\includegraphics[width=0.45\textwidth]{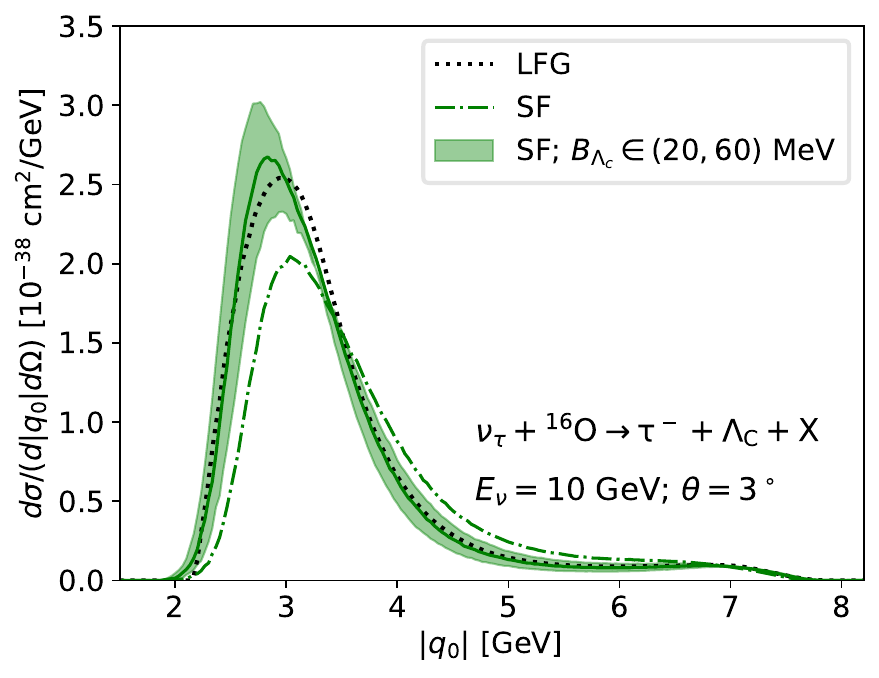}
\includegraphics[width=0.45\textwidth]{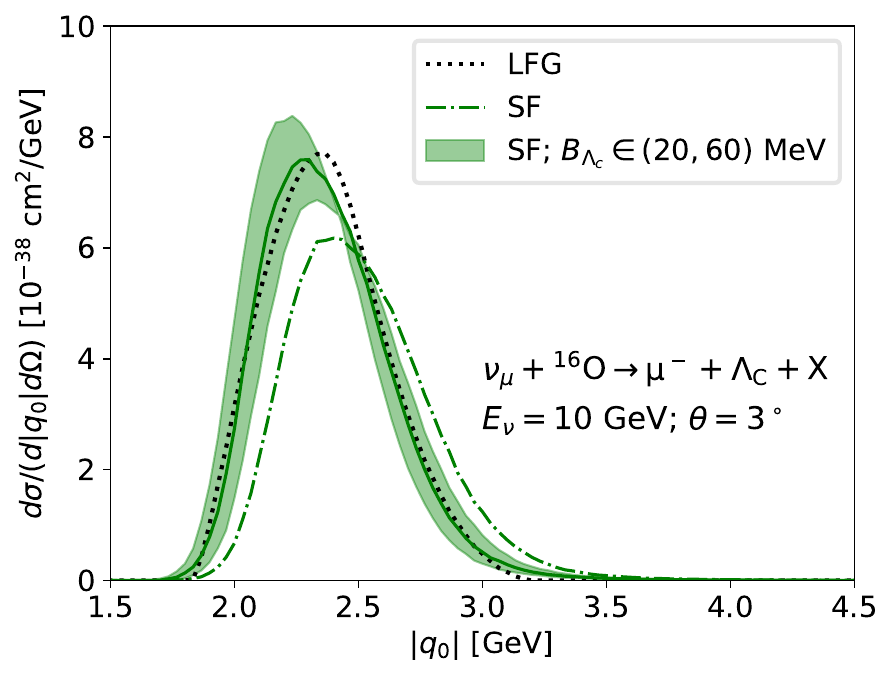}}\\
\makebox[0pt]{\includegraphics[width=0.45\textwidth]{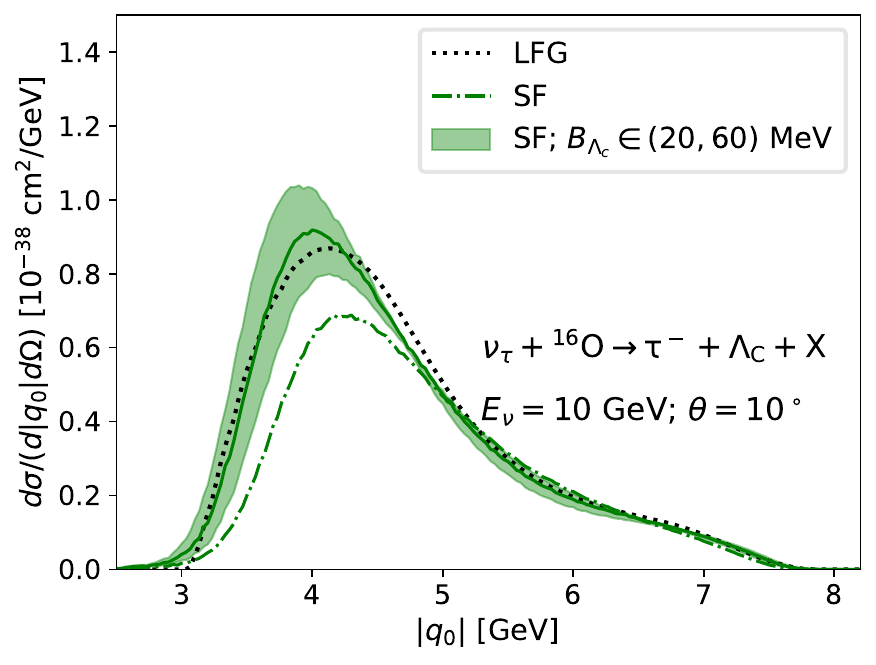}
\includegraphics[width=0.45\textwidth]{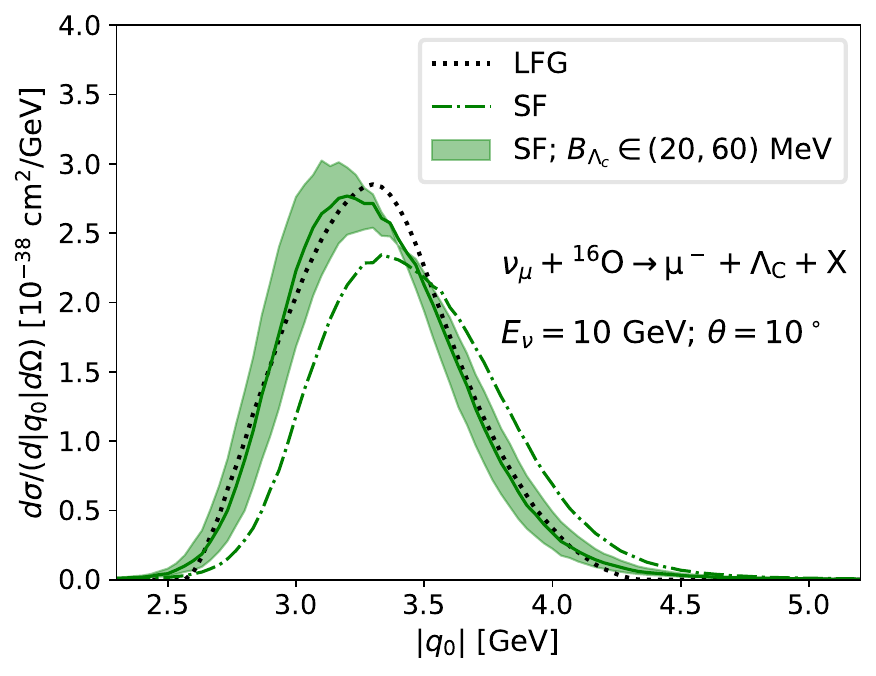}}
\caption{Double-differential cross section  $d^2\sigma/(d\Omega_\ell d|q_0|)$ for  the 
QE  reaction $\nu_\ell+ ^{16}\rm{O}\to \ell^-+\Lambda_c+X$ induced by both tau (left) and muon (right) neutrinos evaluated for $E_\nu=10\,$GeV and $\theta_\ell=3^{\rm o}$ (top) and  $\theta_\ell=10^{\rm o}$ (bottom).} 
\label{fig:res_dq0dOmega_10GeV}
\end{center}
\end{figure}

To get more insight into the role played by nuclear effects, in Fig.~\ref{fig:res_dq0dOmega_10GeV} we present  results for
both tau- and muon-neutrino double differential cross section $d^2\sigma/(d\Omega_\ell d|q^0|)$ on $^{16}$O for $E_\nu=10\,$GeV  and $\theta_\ell=3^{\rm o}$ (top), $\theta_\ell=10^{\rm o}$ (bottom). The use of a realistic hole SF  affects  the total strength, diminishing it by $\sim20\%$ with respect to the LFG calculation. But now  the peak position is also changed. The inclusion of the $\Lambda_c$ binding into the realistic SF calculation enhances the overall strength, bringing it closer to the LFG one, but also moves the peak towards lower values of $|q^0|$. The behavior is similar for both  the $\nu_\tau$ and $\nu_\mu$ induced reactions.  As can be seen in Figs.~\ref{fig:res_dq2_10GeV} and \ref{fig:res_dq0dOmega_10GeV}, the calculation with a realistic SF and a  binding energy $\Lambda_c$ in the  interval $20-60$\,MeV  provides an uncertainty band of $\pm10\%$. This band encompasses, to a reasonable extent, the results obtained with the LFG approach. It provides an estimation of the uncertainties due to our lack of knowledge on the properties of the $\Lambda_c$ baryon embedded in a nuclear medium.

\subsection{Results including BSM physics}

We begin this section by summarizing the present constraints on the relevant WCs, as established in Ref.~\cite{Fuentes-Martin:2020lea} through analysis of $D$-meson decays and high-$p_T$ lepton tails in Drell-Yan data. These constraints were derived under the assumption that only operators involving left-handed neutrinos are present, and that each WC is fitted individually while setting all others to zero.
    
The pure leptonic decay $D^+\to \ell^+\nu_\ell$ with $\ell=e,\mu,\tau$ is able to constrain $1-C^A_{dc\ell L}=C^V_{dc\ell RL}-C^V_{dc\ell LL}$ and $C^P_{dc\ell L}=C^S_{dc\ell LL}-C^S_{dc\ell RL}$. The semileptonic one $D\to \pi^-\ell^+\nu_\ell$, where now $\ell=e,\mu$ (with the $\tau$ decay forbidden by phase space), gives information on $C^V_{dc\ell L}-1=C^V_{dc\ell RL}+C^V_{dc\ell LL}, C^S_{dc\ell L}=C^S_{dc\ell LL}+C^S_{dc\ell RL}$ and $C^T_{dc\ell L}=C^T_{dc\ell LL}$. The limits on each combination of WCs are determined by comparing theoretical results to the average of experimental branching fractions collected by the  Particle Data Group at that time~\cite{Tanabashi:2018oca}.\footnote{See Ref.~\cite{Fuentes-Martin:2020lea} for a full account of the different experimental results included in the averages.}  In Table 2 of Ref.~\cite{Fuentes-Martin:2020lea}, 95\% confidence level (CL) ranges of the above combinations of WCs are collected. We note that in the analysis, they have been further assumed to be real.
        
The analysis of high-$p_T$ lepton tails in Drell-Yan data gives stringent constraints on $C^V_{dc\ell LL}, C^S_{dc\ell LL}, C^S_{dc\ell RL}$ and $C^T_{dc\ell LL}$. Although the contribution from $C^V_{dc\ell LL}$ interferes with the SM one, the results from Ref.~\cite{Fuentes-Martin:2020lea} suggest the term in $|C^V_{dc\ell LL}|^2$ is dominant, with a non-negligible correction coming from  interference with the SM if $C^V_{dc\ell LL}$ is assumed to be real. Note, however, that the interference vanishes exactly if $C^V_{dc\ell LL}$ is purely imaginary. For the others, there is no interference with the SM amplitude and their contributions are already proportional  to the  modulus squared of the corresponding WCs. The 95\% CL bounds are compiled in Table 4 of Ref.~\cite{Fuentes-Martin:2020lea}, with typical magnitudes at the level of $10^{-2}$, in contrast to the SM operator, whose coefficient is of order unity.

The study of Ref.~\cite{Fuentes-Martin:2020lea} was taken in Ref.~\cite{Kong:2023kkd} as the basis to obtain the 1$\sigma$ uncertainties of the  WCs. Then the impact of the different NP terms in the $\nu_\tau n\to\Lambda_c\tau^-$ cross section and  the $\Lambda_c$ and $\tau$ vector polarization components  were  analyzed. The $C^S_{dc\tau LL}, C^S_{dc\tau RL}$ and $C^T_{dc\tau LL}$ WCs are constrained by Drell-Yan data to be of the order of $10^{-2}$. As a result, the influence of the corresponding NP terms in the above observables are simply too small. However, the constraint for $C^V_{dc\tau RL}$, which  results from the $D^+\to\tau^+\nu_\tau$ decay, is given in Ref.~\cite{Kong:2023kkd} at the $1\sigma$ level to be $0.92\lessapprox|1-C^V_{dc\tau RL}|\lessapprox1.16$. This allows for  large values for the real and/or imaginary part of $C^V_{dc\tau RL}$, leading to $\nu_\tau     n\to\Lambda_c \tau$ cross sections  a factor of five larger that SM model ones at high neutrino energies (see Figure 3 in Ref.~\cite{Kong:2023kkd}). This is a very unrealistic scenario and, as suggested in Ref.~\cite{Kong:2023kkd}, one can use this neutrino reaction to put better limits on $C^V_{dc\tau RL}$. Something similar happens for $C^V_{dc\tau LL}$ due to the large imaginary values allowed for this WC in Ref.~\cite{Kong:2023kkd}. As seen in the leftmost panel of their  Figure~2,  there is a large uncertainty interval in the imaginary part of $C^V_{dc\tau LL}$ coming from the leptonic decay. However, as mentioned above, in the case of the  Drell-Yan processes the contribution  from $C^V_{dc\tau LL}$ is  dominated by the quadratic $|C^V_{dc\tau LL}|^2$ term and thus one can consider  the stringent constraint obtained  there  to affect the modulus. This will put a more severe limit on the imaginary part. One would rather expect $|C^V_{dc\tau LL}|\lessapprox10^{-2}$ than having   $|{\rm Im}(C^V_{dc\tau LL})|\lessapprox0.5$ as  assumed in Figure 2 of Ref.~\cite{Kong:2023kkd}. \footnote{Taking $|{\rm Im}(C^V_{dc\tau LL})|\lessapprox0.5$ implies a large deviation from the SM, and it would correspond to the approximate (unrealistic) limit for $|{\rm Im}(C^V_{dc\tau LL})|$ when one assumes that  Drell-Yan data constrain the  real part ${\rm Re}(C^V_{dc\tau LL})$ instead of the modulus of this WC.} With such small values, deviations from the SM, caused by  $C^V_{dc\tau LL}$, for the $\nu_\tau n\to\Lambda_c \tau$ cross section, will also be  minor, contradicting the claims made in Ref.~\cite{Kong:2023kkd}.  This leaves $C^V_{dc\tau RL}$ as the only  WC for which this reaction could be competitive in imposing restrictions on its value.

\begin{figure}[htb]
\begin{center}
\makebox[0pt]{\includegraphics[width=0.45\textwidth]{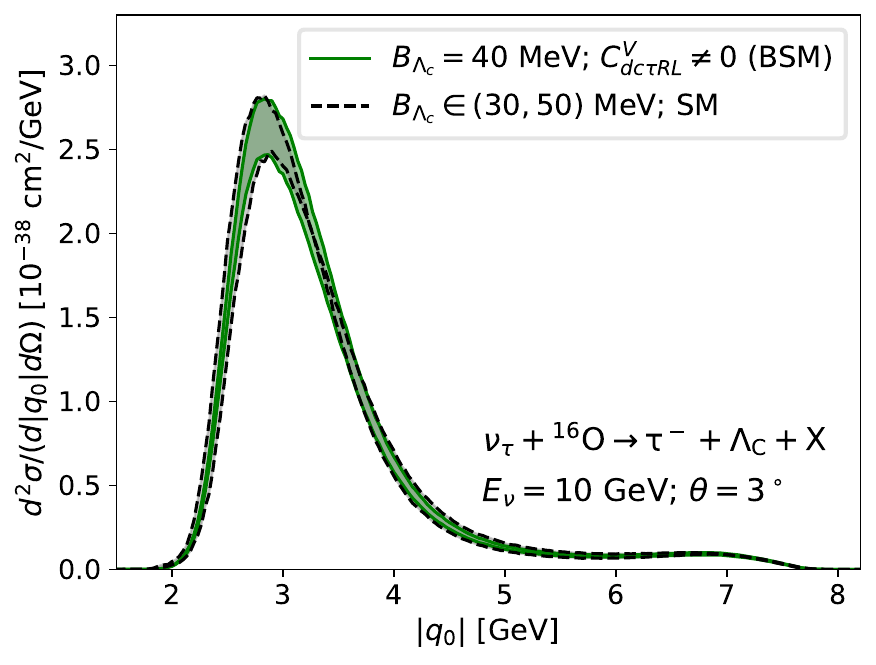}
\includegraphics[width=0.45\textwidth]{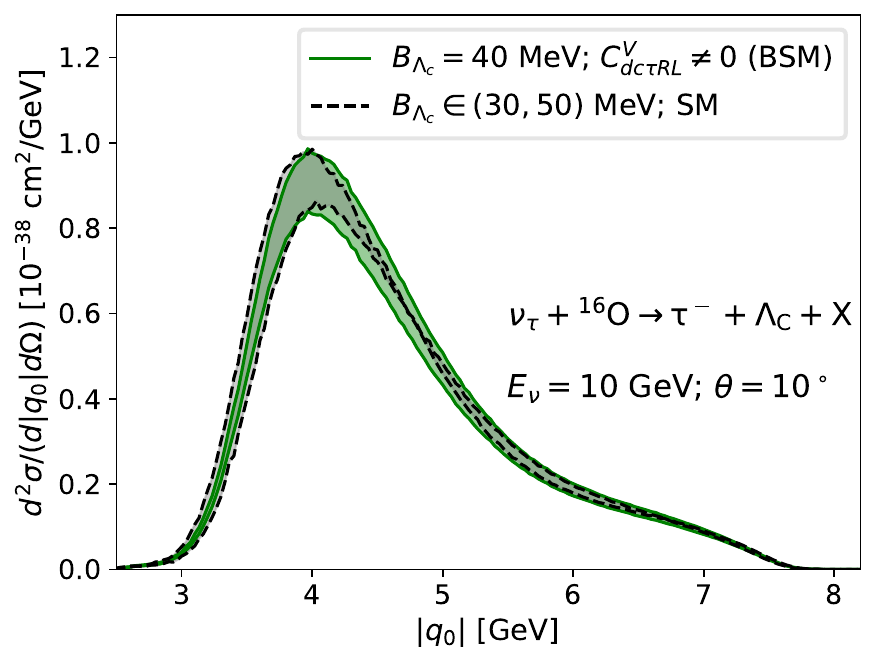}} \\
\makebox[0pt]{\includegraphics[width=0.45\textwidth]{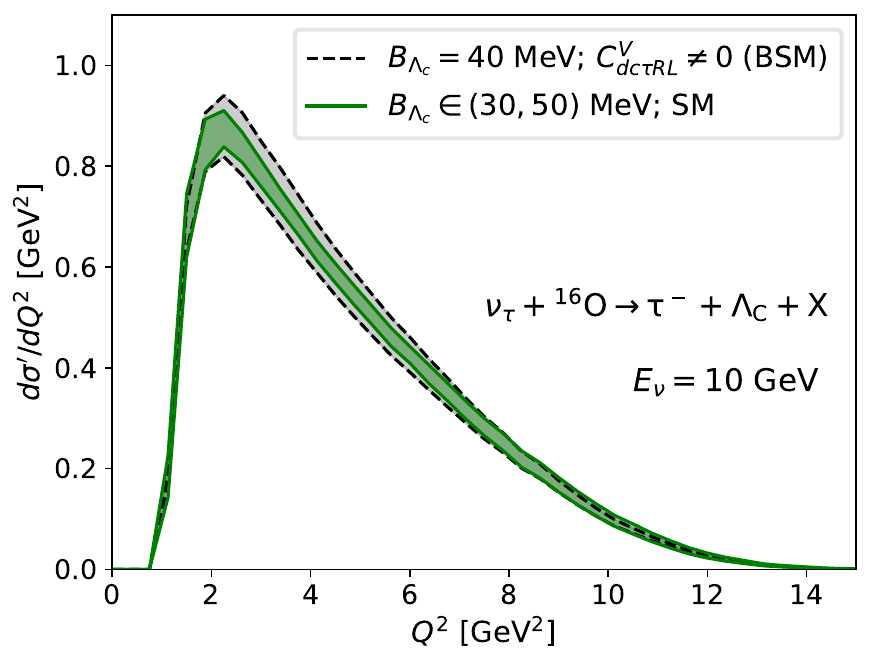}}
\caption{ Nuclear uncertainties versus sensitivity to 
BSM effects.    Top left (right) panel: Double-differential cross section  $d^2\sigma/(d\Omega_\tau d|q_0|)$ for  the QE reaction $\nu_\tau+ ^{16}O\to \tau^-+\Lambda_c+X$  evaluated in  $^{16}$O for $E_\nu=10\,$GeV and $\theta_\tau=3^{\rm o}$ ($\theta_\tau=10^{\rm o}$).  The green band stands for the variation induced in the distribution by the allowed values for the $C^V_{dc\tau RL}$ WC and fixing $B_{\Lambda_c}=40$ MeV.  
The black dashed-lines show the uncertainty in the SM results when $B_\Lambda$ varies in the $30-50$ MeV range. Bottom panel: The $d\sigma'/dQ^2$ [$\sigma'=(8\pi M_n^2\sigma)/(G_F|V_{cd}|)^2$] cross section, per number of neutrons, for the same reaction.}  
\label{fig:BSM1}
\end{center}
\end{figure}

The annular region  $0.92\lessapprox|1-C^V_{dc\tau RL}|\lessapprox1.16$ (see the second panel of Fig.~2 in Ref.~\cite{Kong:2023kkd}), could, in principle, lead to a large deviation from zero in the WC value. Here, however, we adopt a more conservative approach and assume that the SM deviation of this WC should be moderate with $|C^V_{dc\tau RL} - 1|$ not exceeding $\mathcal{O}(10^{-1})$. If the Wilson coefficient is allowed to be complex, this constraint can still accommodate values such that $|C^V_{dc\tau RL} - 0.04| \approx 0.12$.
This variation of $C^V_{dc\tau RL}$ leads only to a moderate modification of the strength of the cross section. The effect, including nuclear corrections, is illustrated in Fig.~\ref{fig:BSM1}.  The nuclear response $F^{A_Z}_{\nu_\tau \Lambda_c}(q)$ has been evaluated following Eq.~\eqref{eq:faz2}, using the nuclear hadron tensor from Eq.\eqref{eq:nucten_SF} and the realistic hole spectral function introduced in Subsec.\ref{sec:realSF}.  The top panels of Fig.~\ref{fig:BSM1} display the double differential cross section $d^2\sigma/(d\Omega_\tau d|q_0|)$ evaluated for $E_\nu=10\,$GeV and the  two  scattering angles, $3^{\rm o}$ and $ 10^{\rm o}$. The lower panel shows $d\sigma/dQ^2$. As indicated by the  green-band, the effect due to the variation of $C^V_{dc\tau RL}$ is of the order of $10\%$ for both observables.
This variation is comparable to the nuclear uncertainty arising from the unknown $\Lambda_c$ binding energy, when the latter is assumed to lie within the range of 30–50 MeV. This nuclear uncertainty is depicted by   the gray-band comprised between the black dashed-lines in Fig.~\ref{fig:BSM1}. As a consequence,  to constrain $C^V_{dc\tau RL}$ via neutrino scattering cross-sections, this level of theoretical precision on nuclear effects would be required. We recall that this analysis is performed under the assumption that the nuclear form factors are known with sufficient accuracy.

\section{Concluding remarks and outlook}
We have obtained completely general expressions for the inclusive (anti-)neutrino-induced QE production of strange and charmed hyperons in nuclei, considering all BSM  dimension-six operators for the semileptonic $q\to q'\ell \nu_\ell$ transition  and both left- and right-handed neutrino fields. 
We illustrated the formalism by applying it to the $\Lambda_c$ production in $\nu_\tau$-nucleus scattering. Our results improve upon those presented in Ref.~\cite{Kong:2023kkd} by explicitly including nuclear corrections and accounting for associated theoretical uncertainties, which were neglected in that earlier work.
At the nucleon level, the overall sizes of the SM cross sections reported in Ref.~\cite{Kong:2023kkd} are a factor of two larger than those found here, a discrepancy we attribute to a likely error in their calculation. Moreover, we argue that the interval  allowed in that work for the imaginary part of the $C^V_{dc\tau LL}$ WC is  inconsistent with the analysis of high-$p_T$ lepton tails in Drell-Yan data  carried out in \cite{Fuentes-Martin:2020lea}. 
To address the nuclear effects, we employed the state-of-the-art hole SF of Ref.~\cite{Sobczyk:2023mey} to model the nuclear ground state, and we analyzed the impact of the final state interactions between the produced hyperon and the residual nuclear system. Our results demonstrate that theoretical nuclear uncertainties may in fact prevent  any NP-discovery claim in this (anti-)neutrino reaction, since their size is  similar, if not much larger, than the signals expected from NP. 

Since the $\nu_\tau(\bar\nu_\tau) A_Z \to \tau^\mp Y X$ reaction is notoriously difficult to be directly measured, the information on the dynamics of this nuclear process, including NP signatures,  should be extracted from the analysis of the energy and angular distributions of the tau decay visible products. These distributions depend, in addition to  $d^{\,2}\sigma/(dE_\tau d\cos\theta_\tau)$,  on the components of the tau-polarization vector. We plan to continue the present research  by obtaining general expressions for the outgoing hadron (charged pion or rho meson) energy and angular differential cross section for the sequential $\nu_\tau A_Z \to \tau^-(\pi^- \nu_\tau, \rho^-\nu_\tau) Y X$ and $\bar\nu_\tau A_Z \to \tau^+(\pi^+ \bar\nu_\tau, \rho^+ \bar\nu_\tau) Y X$ reactions, including NP terms and accounting for nuclear effects. We will consider  the medium corrections by convoluting the single-nucleon cross section and  the hole SF (see Eq.~\eqref{eq:faz1}), since as shown here in Figs.~\ref{fig:res_dq2_10GeV},  \ref{fig:res_2D_10GeV} and \ref{fig:res_dq0dOmega_10GeV}, such procedure reasonably incorporates nuclear effects,  and importantly, will allow us to make use of the exhaustive formalism derived  in Refs.~\cite{Penalva:2021wye} and \cite{Penalva:2022vxy}  for the $H\to H'\tau^-(\pi^- \nu_\tau, \rho^-\nu_\tau)\nu_\tau$ sequential decays of heavy hadrons including general dimension-six BSM operators. 

\section*{Acknowledgments}
This work has been partially supported by MICIU/AEI/10.13039/501100011033
under grants   PID2022-141910NB-I00 and PID2023-147458NB-C21,  by 
Generalitat Valenciana's PROMETEO
program, Ref. CIPROM/2023/59 by the JCyL grant SA091P24 under program 
EDU/841/2024, and by the “Planes Complementarios de I+D+i” program (grant
ASFAE/2022/022) by MICIU with funding from the European
Union NextGenerationEU and Generalitat Valenciana.

\appendix

\section{Relation between the form factors  of Eq.~\eqref{eq.FactoresForma} and the LQCD ones calculated in Ref.~\cite{Meinel:2017ggx} }
\label{app:ff}

One finds 
\begin{eqnarray}
F_1&=&f_{\perp},\quad G_1 =g_{\perp},\quad F_S = \frac{\delta_{M}}{m_d-m_c}f_0,\quad
F_P= \frac{\Delta_{M}}{m_c+m_d}g_0,\quad T_4=\widetilde h_+,\nonumber\\
F_2&=&\frac{M_{n}\delta_{M}}{q^2}f_0+
\frac{M_{n}\Delta_{M}}{s_+}
\left[1-\delta\right]f_+-\delta_{s_{+}}f_{\perp},\nonumber\\
F_3&=&-\frac{M_{\Lambda_c}\delta_{M}}{q^2}f_0
+\frac{M_{\Lambda_c}\Delta_{M}}{s_+}
\left[1+\delta\right]f_+-\delta_{s_{+}}f_{\perp},\nonumber \\
G_2&=&-\frac{M_{n}\Delta_{M}}{q^2}g_0-\frac{M_{n}\delta_{M}}{s_-}\left[1-\delta\right]g_+
-\delta_{s_{-}}g_{\perp},\nonumber\\
G_3&=&\frac{M_{\Lambda_c}\Delta_{M}}{q^2}g_0-\frac{M_{\Lambda_c}
\delta_{M}}{s_-}\left[1+\delta\right]g_+
+\delta_{s_{-}}g_{\perp},\nonumber\\
\nonumber\\
T_1&=&-2M^2_{n}\Big(\frac{h_+}{s_+}-\frac{\Delta^2_M}{q^2s_+}h_\perp-
\frac{\tilde h_+}{s_-}+\frac{\delta^2_M}{q^2s_-}\,\tilde h_\perp\Big),\nonumber\\
T_2&=&M_n\Big(\frac{\Delta_M}{q^2}h_\perp+
\frac{2M_{\Lambda_c}}{s_-}\tilde
h_+ +\frac{\delta_{M}}{s_-}[1-\delta]\,\tilde h_\perp
\Big),\nonumber\\
T_3&=&M_n\Big(-\frac{\Delta_M}{q^2}h_\perp-
\frac{2M_{n}}{s_-}\tilde h_++\frac{\delta_{M}}{s_-}[1+\delta]\,\tilde h_\perp
\Big).
\end{eqnarray}
with $\delta=(M_{n}^2-M_{\Lambda_c}^2)/q^2$, 
$s_{\pm}=(M_{\Lambda_c}\pm M_{n})^2-q^2$, 
$\delta_{M}=M_{n}-M_{\Lambda_c}$, 
$\Delta_{M}=M_{\Lambda_c}+M_{n}$ and $\delta_{s_{\pm}}=
2M_{\Lambda_c}M_{n}/s_{\pm}$.
Note that $F_S$ and $F_P$ have not been computed in LQCD, and both form factors 
are obtained from the vector $f_0$ and axial $g_0$ form factors using the 
equations of motion. In the numerical calculations, we use   
$m_c = 1.2730 \pm 0.03046$\,GeV and $m_d=4.77\pm0.07$\,MeV taken from 
Ref.~\cite{ParticleDataGroup:2024cfk}.

{The expressions above agree with the ones in appendix E of 
Ref.~\cite{Penalva:2020xup} with the replacements $M_{\Lambda_b} \to M_n$ 
and $m_b \to m_d$. }

\section{Evaluation of $F^{A_Z}_{\nu_\tau \Lambda_c}(q)$ in Eq.~\eqref{eq:faz1}}
\label{app:a012b01c0d0}
We first use   rotational covariance  and introduce the functions
\bea
\int\frac{d^3p}{(2\pi)^3}\int dES_h^{(n)}(E,|\vec p\,|)
\frac{M_nM_{\Lambda_c}}{\sqrt{M_n^2+\vec{p}^{\,2}}\, E'}\,
 \delta(E-E'-q^0)\,E^m &=&a_m, \qquad m=0,1,2 \label{eq:am}\\
\int\frac{d^3p}{(2\pi)^3}\int dES_h^{(n)}(E,|\vec p\,|)
\frac{M_nM_{\Lambda_c}}{\sqrt{M_n^2+\vec{p}^{\,2}}\, E'}\,
 \delta(E-E'-q^0)\,E^mp^j &=&
b_m\, q^j, \qquad m=0,1  \label{eq:bm}\\
\int\frac{d^3p}{(2\pi)^3}\int dES_h^{(n)}(E,|\vec p\,|)
\frac{M_nM_{\Lambda_c}}{\sqrt{M_n^2+\vec{p}^{\,2}}\, E'}\,
 \delta(E-E'-q^0)\,p^jp^k &=&
c_0|\vec q\,|^2\, \delta^{jk}+d_0\,q^jq^k.
\label{eq:c0d0}
\eea
They depend on $q^0$ and $|\vec q\,|$ and   can be evaluated
once the neutron-hole SF is known.\footnote{Note that in the case of a
 $d\to u$ driven transition producing a $p$ instead of $\Lambda_c^+$ hyperon, 
 in the above integrations one should, at least,  account for the Pauli 
 blocking of the final proton. }.  
 
Recalling that $ E'=\sqrt{M^2_{\Lambda_c}+(\vec
p-\vec q\,)^2}$, $q=k'-k$,  it trivially follows 
\bea
a_m&=&\frac{M_n M_{\Lambda_c}}{4\pi^2|\vec q\,|}\int  d|\vec p\,||\vec p\,|\int
\frac{dE \,S_h^{(n)}(E,|\vec p\,|)}{\sqrt{M_n^2+\vec{p}^{\,2}}} E^m H\left[1-|x_0|\right], \quad m=0,1,2\no
b_m&=&\frac{M_n M_{\Lambda_c}}{4\pi^2|\vec q\,|^2}\int 
d|\vec p\,||\vec p\,|^2 \int
\frac{dE \,S_h^{(n)}(E,|\vec p\,|)}{\sqrt{M_n^2+\vec{p}^{\,2}}} E^m\,x_0\,H\left[1-|x_0|\right], \quad m=0,1\no
c_0&=&\frac{M_n M_{\Lambda_c}}{8\pi^2|\vec q\,|^3}
\int  d|\vec p\,||\vec p\,|^3\int
\frac{dE \,S_h^{(n)}(E,|\vec p\,|)}{\sqrt{M_n^2+\vec{p}^{\,2}}}(1-x_0^2)H\left[1-|x_0|\right],\no
d_0&=&\frac{M_n M_{\Lambda_c}}{8\pi^2|\vec q\,|^3}
\int  d|\vec p\,||\vec p\,|^3\int
\frac{dE \,S_h^{(n)}(E,|\vec p\,|)}{\sqrt{M_n^2+\vec{p}^{\,2}}}(-1+3x_0^2)H\left[1-|x_0|\right].
\eea
with $x_0=\left(M^2_{\Lambda_c}+|\vec p\,|^2-E^2-q^2+2q^0 E\right)/\left(2|\vec p\,|\,|\vec
q\,|\right)$.\footnote{This is modified to $x_0=\left(M^2_{\Lambda_c}+|\vec p\,|^2-E^2-\hat q^2+2\hat q^0 E\right)/\left(2|\vec p\,|\,|\vec
q\,|\right)$, with $\hat q=(q^0-B_{\Lambda_c},\vec q\,)$, if the binding for the $\Lambda_c$ is included.} As we see all angular integrations have been done analytically which reduces the uncertainties.

In terms of these functions one has that $F^{A_Z}_{\nu_\tau \Lambda_c}(q)$ defined in Eq.~\eqref{eq:faz1}
is given by
\bea
 F^{A_Z}_{\nu_\tau \Lambda_c}(q) &=&
\frac{M^2_n }{2}\Bigg\{
\frac{{\cal B}(\omega)}{M^2_n}\left(a_1|\vec k\,|-b_0\, \vec q\cdot\vec k\right)-{\cal A}(\omega)a_0-
\frac{{\cal C}(\omega)}{M^4_n}\left( |\vec k\,|^2a_2+c_0|\vec q\,|^2|\vec k\,|^2+ d_0\,(\vec q\cdot\vec k)^2-
2b_1|\vec k\,|\,\vec q\cdot\vec k\right) \label{eq:nucl-ampl}
\Bigg\}.\no
\eea

For the case of a LFG,  using the corresponding  hole SF of Eq.~\eqref{eq:ShLFG} and no extra $B_{\Lambda_c}$ binding,\footnote{Remember that, for a LFG,  we are implicitly  assuming a local $\Lambda_c$ binding equal to the one of the nucleon.} one further has that the $a_m$ defined above
are given by
\be
a^{\rm LFG}_m =\frac{2M_n M_{\Lambda_c}}{(m+1)\pi|\vec q\,|}\int r^2dr\, H\left [E_F(r)-\tilde E\right]\left(E^{m+1}_F(r)-\tilde
E^{m+1}\right), \quad m=0,1,2 \label{eq:aLFGm}
\ee
where we have used that in the LFG case, neglecting off-shell effects, 
\bea
1-x_0^2&=&-\frac{q^2}{|\vec q\,|^2|\vec p\,|^2}\left(E-{\cal E}_+\right)\left(E-{\cal E}_-\right) \label{eq:x0}\\
{\cal E}_{\pm}&=&\frac{q^0}{2}\left(1-\frac{M^2_{\Lambda_c}-M^2_n}{q^2}\right)
\pm\frac{|\vec q\,|}{2}\sqrt{\left(1-\frac{M^2_{\Lambda_c}-M^2_n}{q^2}\right)^2-\frac{4M^2_n}{q^2}}
\eea
where ${\cal E}_-< 0$ (since $q^2<0$), and  $ {\cal E}_+>0$ is the minimum neutron energy that guarantees that $x_0\le 1$. In addition in Eq.~\eqref{eq:aLFGm}, $E_F(r)=\sqrt{M^2_n+k^2_F(r)}$ and $\tilde E=\max\{M_n,{\cal E}_+\}$.
Besides, since the neutron is on-shell, we have that  
$q^0 E-\vec p\cdot\vec q=(M^2_n+q^2-M^2_{\Lambda_c})/2$, and from
Eqs.~\eqref{eq:am}-\eqref{eq:c0d0}  it readily follows
\be
b^{\rm LFG}_{m-1}=a^{\rm LFG}_m \frac{q^0}{|\vec
q\,|^2 }-\frac{a^{\rm LFG}_{m-1}}{2|\vec
q\,|^2 }\left(M^2_n+q^2-M^2_{\Lambda_c}\right), \qquad m=1,2
\ee
and 
\bea
c^{\rm LFG}_0 &=& -\frac{1}{2|\vec
q\,|^4}\left\{q^2a^{\rm LFG}_2 - q^0 a^{\rm LFG}_1  \left(q^2-M^2_{\Lambda_c}+M^2_n\right) +\frac{a^{\rm LFG}_0}{4}\left[ \left(q^2-M^2_{\Lambda_c}+M^2_n\right)^2+ 4M_n^2|\vec
q\,|^2\right]  \right\}\\
d^{\rm LFG}_0 &=& -3c^{\rm LFG}_0  + \frac{a^{\rm LFG}_2-M^2_na^{\rm LFG}_0}{|\vec
q\,|^2}
\eea
In the case of the $d\to u$ driven transition producing 
a $p$ instead of $\Lambda_c^+$ hyperon, one should account for the Pauli 
blocking of the final proton, replacing the $\delta[E'-(E-q^0)]$ in 
Eqs.~\eqref{eq:am}-- \eqref{eq:c0d0} by a particle spectral 
function~\cite{Nieves:2017lij}. In that case, and for a LFG, all integrals in 
Eqs.~\eqref{eq:am}-- \eqref{eq:c0d0} will be related to the imaginary part 
of the Lindhard function (see Appendix B of Ref.~\cite{Nieves:2004wx}).

\bibliography{LCTAU}
\end{document}